%% file: main-2.tex
\documentclass[aps,prb,amsmath,amssymb,superscriptaddress,reprint,floatfix]{revtex4-2}

\usepackage{cmap}
\usepackage[T1]{fontenc}
\usepackage{newtxtext}
\usepackage[varvw]{newtxmath}
\usepackage{microtype}
\usepackage{graphicx}
\usepackage{bm}
\usepackage{hyperref}
\hypersetup{colorlinks=true,allcolors=blue}
\usepackage{mathtools}
\usepackage{array}
\usepackage{upgreek}
\usepackage{orcidlink}
\usepackage{makecell}

\renewcommand{\leq}{\leqslant}

\usepackage{color}
\usepackage{physics, ulem}
\usepackage{cases}
\usepackage{mathrsfs}
\usepackage{subfigure}
\usepackage{comment}
\usepackage{physics2}
\usephysicsmodule{ab}

\newcommand{\del}{\partial}

\newcommand{\sgn}{\mathrm{sgn}}

\newcommand{\lb}{\left(}
\newcommand{\rb}{\right)}

\newcommand{\LB}{\left[}
\newcommand{\RB}{\right]}
\newcommand{\Lb}{\langle}
\newcommand{\Rb}{\rangle}

\newcommand{\up}{\uparrow}
\newcommand{\down}{\downarrow}

\begin{document}

\title{Charge-spin conversion in altermagnets: electronic anisotropy versus magnon-mediated spin drag}

\author{Koki Mizuno\,\orcidlink{0009-0004-8276-1045}}
\email{mizuno.koki.t8@s.mail.nagoya-u.ac.jp}
\affiliation{
    Department of Physics,
    \href{https://ror.org/04chrp450}{Nagoya University},
    Nagoya, Japan
}

\begin{abstract}
    Altermagnets acquire a momentum-dependent spin splitting without net magnetization or spin-orbit coupling.
    This raises the question of whether the transport anisotropy originates in the electrons or the magnons, and whether the magnon-mediated channel distinguishes an altermagnet from a conventional antiferromagnet.
    We compute the spin-resolved conductivity of a metallic ferromagnet, antiferromagnet, and altermagnet within a common model, treating the spin-diagonal channel with the magnon self-energy and the spin-flip channel to leading order in the electron-magnon coupling.
    The anisotropy of the charge-spin conversion is dominated by the electronic spin splitting, the magnon branches contributing only marginally.
    The magnon-mediated spin-flip channel is even under $C_{4}$, so it carries no signature at first order in the anisotropy and cannot on its own distinguish an altermagnet from a conventional antiferromagnet.
    It does, however, respond linearly to a sublattice-selective perturbation that lowers the bulk $B_{1g}$ symmetry.
    The resulting susceptibility vanishes identically in the antiferromagnet, is small in the ferromagnet, and is finite in the altermagnet, where the electronic anisotropy ties spin and sublattice together, so that among magnets with no net magnetization it is unique to altermagnetic order.
    Being even in the magnetic domain index, this response can be measured without preparing a single magnetic-domain sample.
\end{abstract}

\maketitle

\section{Introduction}
Altermagnetism has recently been recognized as a third class of collinear magnetic order, distinct from both ferromagnetism and conventional antiferromagnetism, and characterized by momentum-dependent spin splitting of nonrelativistic origin \cite{SpinSplitting_2019_Hayami,SpinSplitting_2020_Yuan,SpinSplitting_2020_Hayami,SpinSplitting_2021_Yuan,Altermagnet_2022_Smejkal-PRX031042,Altermagnet_2022_Smejkal-PRX040501,Altermagnet_spintronics_2025_Song,Altermagnet_spintronics_2025_Tamang}.
This nonrelativistic spin splitting is classified by spin-group symmetry \cite{Altermagnet_2022_Liu}, which distinguishes altermagnets from ferromagnets and conventional antiferromagnets.
The spin splitting is predicted to generate a range of spin-dependent transport phenomena in symmetry-allowed altermagnets, including the spin-splitter effect, without requiring either spin-orbit coupling or net magnetization \cite{Cite-in-Almag-S-E-conv_2019_Naka,SpinCurrent_2021_Naka,SpinCurrent_2021_Ma,SpinSplitter_2021_Gonzalez-Hernandez,SpinNeutral_2021_Shao,Magnetoresistance_2022_Smejkal,NeelSpinCurrent_2023_Shao,Cite-in-Almag-S-E-conv_2025_Dou,ChargeSpinConversion_2025_Lai,Cite-in-Almag-S-E-conv_2026_Sigales,SpinCurrent_2026_Sheoran}.
This prediction has been tested in RuO$_2$/ferromagnet bilayers, where spin-splitting torques and direct or inverse spin--charge conversion signals with characteristic dependencies on the crystal axes and the N\'eel-vector orientation have been reported \cite{Cite-in-Almag-S-E-conv_2022_Bai,Cite-in-Almag-S-E-conv_2022_Karube,Cite-in-Almag-S-E-conv_2022_Bose,SpinChargeConversion_2023_Bai,DirectInverseSSE_2024_Guo,InverseASSE_2024_Liao,Altermagnet_spintronics_2025_Zhang}, although both the magnetic order of RuO$_2$ and the attribution of its transport signals to altermagnetism have since been called into question \cite{Cite-in-Almag-S-E-conv_2024_Hiraishi,Cite-in-Almag-S-E-conv_2024_Kessler,AbsenceSpinSplitting_2024_Liu,AbsenceTransportASSE_2026_Wang}.

Altermagnets can likewise host symmetry-allowed splitting of oppositely polarized magnon modes \cite{ChiralMagnon_2023_Smejkal,ChiralMagnon_2024_Liu,Cite-in-Almag-S-E-conv_2025_Garcia-Gaitan,ChiralMagnon_2025_Kravchuk,Cite-in-Almag-S-E-conv_2025_Beida,ChiralMagnonRIXS_2025_Biniskos,Cite-in-Almag-S-E-conv_2025_Sun,Cite-in-Almag-S-E-conv_2026_Liu}, although the splitting can be negligible or dominated by dipolar interactions \cite{AbsenceChiralMagnon_2025_Morano,MagnonSplitting_2026_Sears}.
Consequent phenomena include anisotropic magnon transport, magnon drag, and spin pumping \cite{MagnonSpinSeebeckNernst_2023_Cui,MagnonSpinSeebeckNernst_2024_Weissenhofer,AltermagnonDrag_2025_Sourounis,SpinPumping_2024_Hodt,MagnonInterference_2025_Sheng,MagnonOrbitalNernst_2026_Weissenhofer,Cite-in-Almag-S-E-conv_2026_Lei}.
Electron--magnon coupling also modifies electron and magnon lifetimes \cite{QuasiparticleLifetime_2026_Leraand,Cite-in-Almag-S-E-conv_2025_Weber,Cite-in-Almag-S-E-conv_2025_Beida} and can mediate superconductivity \cite{MagnonSC_2024_Maeland,Cite-in-Almag-S-E-conv_2023_Brekke}.
Yet whether magnon-mediated electronic transport can differentiate an altermagnet from a conventional antiferromagnet remains unclear.

In this paper, we construct the charge-spin conversion matrix from the spin-resolved longitudinal conductivity tensor of a metallic ferromagnet, antiferromagnet, and altermagnet, evaluating the spin-diagonal channel at one-loop order and the magnon-mediated spin-flip channel at leading order in the electron-magnon coupling.
The charge-spin conversion, carried by the spin-diagonal channel, retains the same $d$-wave anisotropy as the electronic band structure, inherited from the same nonrelativistic spin splitting that underlies the spin-splitter effect, and the magnon anisotropy gives only a minor correction.
In contrast, the magnon-mediated spin-flip channel, isolated as the difference between the electronic and the spin conductivity, is even under $C_4$ rotation.
As a result, the antiferromagnet and the altermagnet behave almost identically, and the altermagnetic order leaves no signature at first order in the electronic and magnonic anisotropy that distinguishes it from the antiferromagnet.
Our central result concerns the response of this channel to a perturbation that lowers the bulk $B_{1g}$ symmetry, implemented here as a sublattice-selective modulation of the electron-magnon coupling.
The resulting susceptibility vanishes identically in the antiferromagnet, is small in the ferromagnet, and is finite in the altermagnet, where the electronic anisotropy ties the spin and the sublattice degrees of freedom together.
The vanishing in the antiferromagnet is exact and follows from symmetry alone, whereas the separation from the ferromagnet is quantitative and holds in the low frequency region.
Since a ferromagnet is identified by its net magnetization in any case, a finite susceptibility at low frequency is, among magnets with no net magnetization, unique to altermagnetic order.
Decomposing it into its electronic and magnonic sources shows that the magnonic contribution is small and of the opposite sign, so that the anisotropy relative to the isotropic part is governed by the electronic anisotropy alone.
Since this susceptibility is even under a magnetic domain reversal, it can be measured without preparing a single magnetic-domain sample.

The rest of this paper is organized as follows.
Section~\ref{sec:model} introduces the model and the propagators, Sec.~\ref{sec:matrix} defines the charge-spin conversion matrix and derives the spin-diagonal and spin-flip conductivities together with their behavior under a domain reversal, and Sec.~\ref{sec:results} presents the numerical results.

\section{Model and Green functions}\label{sec:model}
Unless otherwise stated, we use units in which $k_{\rm B} = e = \hbar=1$; physical constants are restored explicitly when needed.

\begin{figure*}
    \centering 
    \includegraphics[scale=0.7]{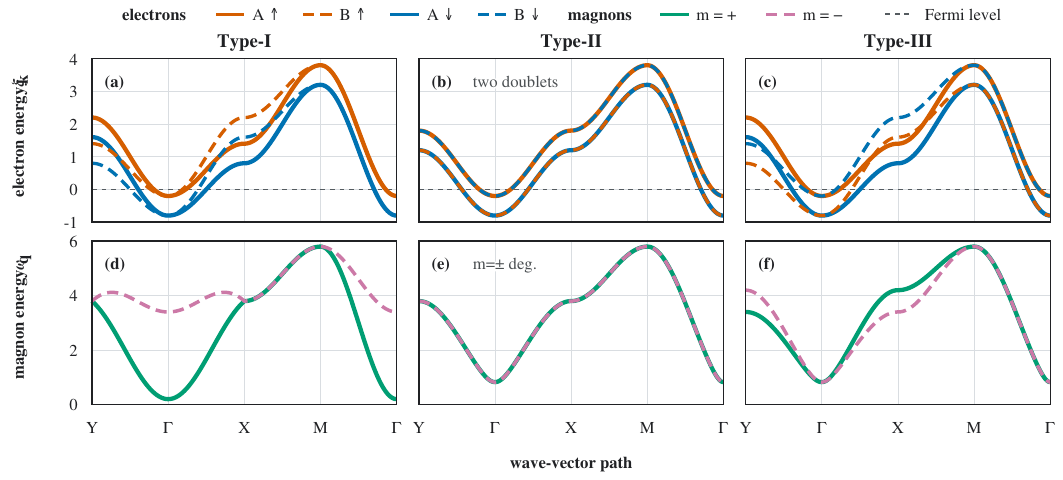}
    \caption{
        Schematic of the band structure for the type-I, type-II, and type-III magnets.
        The upper panels (a-c) show the electronic band structure, and the lower panels (d-f) show the magnon band structure.
        For the electronic band, the solid and dashed lines represent the sublattice $A$ and $B$, respectively, and the orange and blue lines represent the spin $\up$ and $\down$, respectively.
        For the magnon band, the solid and dashed lines represent the magnon mode $m = +1$ and $-1$, respectively.
        The magnetic types I, II, and III are shown in the left, middle, and right panels, respectively.
        In this figure, we set the magnetic domain $\mathcal{N} = +1$, temperature $T = 0.2$, the chemical potential $\mu = 0.5$, and the other parameters are set as $t = J = S = a = 1$, $t' = 0.2$, $\Delta = M = 0.3$, $J_{0} = 0.4$, $K=0.1$ and $\delta J = 0.2$.
    }
    \label{fig:energy}
\end{figure*}

We use a minimal four-band model allowed by the $D_{4h}$ symmetry of the square lattice to compare the type-I, type-II, and type-III magnets.
The Hamiltonians are given by
\begin{align}
    H_{\rm I}(\vv{k}) &= \epsilon_{0}(\vv{k}) I_{2}\otimes I_{2} + d_{z}(\vv{k}) I_{2} \otimes \tau_{z} + \Delta \sigma_{z} \otimes I_{2},
    \\
    H_{\rm II}(\vv{k}) &= \epsilon_{0}(\vv{k}) I_{2}\otimes I_{2} + M \sigma_{z} \otimes \tau_{z}, 
    \\
    H_{\rm III}(\vv{k}) &= \epsilon_{0}(\vv{k}) I_{2}\otimes I_{2} + d_{z}(\vv{k}) I_{2} \otimes \tau_{z} + M \sigma_{z} \otimes \tau_{z},
\end{align}
where $\sigma_{x,y,z}$ and $\tau_{x,y,z}$ are the Pauli matrices acting in the spin and $A/B$ sublattice spaces, respectively.
The dispersions are $\epsilon_{0}(\vv{k}) = t \lb 2 - \cos(a k_{x}) - \cos(a k_{y}) \rb$ and $d_{z}(\vv{k}) = t' \lb \cos(a k_{x}) - \cos(a k_{y}) \rb$.
We define $\eta_A=+1$ and $\eta_B=-1$ as the eigenvalues of $\tau_z$, and use $\sigma=+1$ ($-1$) for spin $\up$ ($\down$).
For a magnetic domain $\mathcal{N}=\pm1$, the band energies are
\begin{align*}
    \epsilon_{\vv{k}\alpha\sigma}^{\rm I}
    &= \epsilon_{0}(\vv{k}) + \eta_{\alpha}d_{z}(\vv{k}) + \mathcal{N}\sigma\Delta,
    \\
    \epsilon_{\vv{k}\alpha\sigma}^{\rm II}
    &= \epsilon_{0}(\vv{k}) + \mathcal{N}\sigma\eta_{\alpha}M,
    \\
    \epsilon_{\vv{k}\alpha\sigma}^{\rm III}
    &= \epsilon_{0}(\vv{k}) + \eta_{\alpha}d_{z}(\vv{k}) + \mathcal{N}\sigma\eta_{\alpha}M.
\end{align*}
Here, $\mathcal{N}=+1$ corresponds to the Hamiltonians displayed above, while $\mathcal{N}=-1$ denotes their time-reversed magnetic domain.
One can verify that these models satisfy the constraints of the spin point group with the sublattice-space representation given by $D_{\tau}(g) = I_{2}$ for $g \in D_{2h}$ and $D_{\tau}(g) = \tau_{x}$ for $g \in (D_{4h} - D_{2h})$.
The structure of the band is shown in Fig.~\ref{fig:energy}.
The type-I magnet (a) has the isotropic spin splitting, the type-II magnet (b) has the spin-degenerate bands, and the type-III magnet (c) has the $d$-wave spin splitting.
Note that the type-I magnet is written with two sublattices, and retains the sublattice-staggered term $d_{z}(\vv{k})\tau_{z}$, so that the three magnetic orders can be compared within a common four-band model; a ferromagnet with a single site per unit cell would not have this term.

The corresponding localized-spin systems can be described by Heisenberg
models on the same square lattice.
We denote the position of a site on sublattice $\alpha$ by $\vv{r}_{\alpha}$.
The four nearest-neighbor bond vectors from an $A$ site to the surrounding
$B$ sites are
\begin{align}
    \mathcal{D}_{AB}
    = \left\{
    \frac{a}{2}\left(\eta_x\hat{\vv{x}}+\eta_y\hat{\vv{y}}\right)
    \;\middle|\; \eta_x,\eta_y=\pm1
    \right\}.
\end{align}
The corresponding bond structure factor is
$\Gamma_{\vv{q}}=\sum_{\vv{\delta}\in\mathcal{D}_{AB}}
e^{i\vv{q}\cdot\vv{\delta}}
=4\cos(aq_x/2)\cos(aq_y/2)$.
The sums over $\vv{r}_A$ and $\vv{\delta}$ below count each
intersublattice bond once. For the same-sublattice bonds, we use the
normalization in which their exchange coefficients are $J/2$ and
$J_{\mu}^{\alpha}/2$, respectively; with this convention, $J$ and
$J_{\mu}^{\alpha}$ are the parameters entering the magnon dispersions below.
For a type-I magnet, we introduce two symmetry-related sublattices, $A$ and $B$,
in each unit cell and couple their spins ferromagnetically,
\begin{align}
    \mathcal{H}_{\rm I}^{\rm spin}
    ={}&-J_{0}\sum_{\vv{r}_A}
    \sum_{\vv{\delta}\in\mathcal{D}_{AB}}
    \vv{S}_{\vv{r}_A, A}\cdot
    \vv{S}_{\vv{r}_A+\vv{\delta}, B}
    \notag\\
    &-\frac{J}{2}\sum_{\vv{r},\mu=x,y}\sum_{\alpha=A,B}
    \vv{S}_{\vv{r}_{\alpha},\alpha}\cdot
    \vv{S}_{\vv{r}_{\alpha}+a\hat{\vv{\mu}},\alpha}
    \notag\\
    &-K\sum_{\vv{r},\alpha}
    \left(S_{\vv{r}_{\alpha},\alpha}^{z}\right)^{2},
    \qquad J_{0},J>0 ,
    \label{eq:spin_type_I}
\end{align}
where the easy-axis anisotropy $K>0$ selects the collinear direction.
The classical ground state is the ferromagnetic state
$\vv{S}_{\vv{r}_A, A}=\vv{S}_{\vv{r}_B, B}=S\hat{\vv{z}}$, which is the
spin-system counterpart of the sublattice-independent exchange field
$\Delta\sigma_z\otimes I_2$ in $H_{\rm I}$.
As shown in Appendix~\ref{app:FM_magnon}, the Holstein-Primakoff transformation describes the ferromagnetic magnon excitations around the collinear ground state.
The magnon Hamiltonian is written as
\begin{align}
    \mathcal{H}_{\rm I}^{\rm mag}
    =\sum_{\vv{q},m=\pm}
    \left(\gamma_{\vv{q}} + m \gamma_{0,\vv{q}}^{\rm I}\right)
    a_{\vv{q}m}^\dagger a_{\vv{q}m},
\end{align}
where $\gamma_{0,\vv{q}}^{\rm I} = -J_{0}S\Gamma_{\vv{q}}$, $\gamma_{\vv{q}} = (4J_{0} + 2K)S + JS \sum_{\mu} \left(1- \cos(a q_{\mu})\right)$, and $a_{\vv{q}\pm}$ is the magnon annihilation operator.
This energy spectrum is shown in Fig.~\ref{fig:energy}(d). At the $\Gamma$ point, $K$ opens the lower-branch gap $2KS$, while the intersublattice exchange $J_{0}$ gives the branch separation $8J_{0}S$.
The exchange interaction between the spin and electron systems is given by
\begin{align}
    \mathcal{H}_{\rm ex}
    =-J_{\rm ex}\sum_{\vv{r},\alpha}
    \vv{S}_{\vv{r}_{\alpha},\alpha}\cdot
    \vv{s}_{\vv{r}_{\alpha},\alpha},
    \label{eq:exchange_interaction}
\end{align}
where $\vv{s}_{\vv{r}_{\alpha},\alpha} = \frac{1}{2}\sum_{\sigma,\sigma'} c_{\vv{r}_{\alpha},\alpha\sigma}^\dagger\vv{\sigma}_{\sigma\sigma'} c_{\vv{r}_{\alpha},\alpha\sigma'}$ is the electron spin operator at position $\vv{r}_{\alpha}$ on sublattice $\alpha$.
Retaining terms through first order in the magnon operators and neglecting the subleading longitudinal term $J_{\rm ex}\sum_{\vv{r},\alpha}a_{\vv{r}_{\alpha},\alpha}^{\dagger}a_{\vv{r}_{\alpha},\alpha}s_{\vv{r}_{\alpha},\alpha}^{z}$, we obtain
\begin{align}
    \mathcal{H}_{\rm ex}
    \simeq{}&-J_{\rm ex}S\sum_{\vv{r},\alpha} s_{\vv{r}_{\alpha},\alpha}^{z}
    \notag\\
    &-\frac{J_{\rm ex}\sqrt{2S}}{2}\sum_{\vv{r},\alpha}
    \left(a_{\vv{r}_{\alpha},\alpha}^\dagger s_{\vv{r}_{\alpha},\alpha}^{+} + a_{\vv{r}_{\alpha},\alpha} s_{\vv{r}_{\alpha},\alpha}^{-}\right).
\end{align}
The first term corresponds to the exchange field $\Delta\sigma_z\otimes I_2$ in $H_{\rm I}$, and we set $\Delta = -J_{\rm ex}S/2$.
Thus, the magnon creating and annihilating interaction in the Fourier space is written as
\begin{align}
    \mathcal{H}_{I,\rm int}
    &= \sqrt{\frac{2}{S}} \Delta \sum_{\vv{q},\alpha}
    \left(a_{\vv{q}\alpha}^\dagger s_{-\vv{q},\alpha}^{+} + a_{\vv{q}\alpha} s_{\vv{q},\alpha}^{-}\right),
    \notag\\ 
    &=
    \frac{\Delta}{\sqrt{S}} \sum_{\vv{q}}
    \left\{
        \lb s_{-\vv{q},A}^{+} + s_{-\vv{q},B}^{+} \rb a_{\vv{q}+}^\dagger  
        + \lb s_{\vv{q},A}^{-} + s_{\vv{q},B}^{-} \rb a_{\vv{q}+}
    \right.
    \notag\\
    &\qquad\quad
    \left.
        + \lb s_{-\vv{q},A}^{+} - s_{-\vv{q},B}^{+} \rb a_{\vv{q}-}^\dagger
        + \lb s_{\vv{q},A}^{-} - s_{\vv{q},B}^{-} \rb a_{\vv{q}-}
    \right\}.
\end{align}
The Fourier transforms are
\begin{align*}
    s_{\vv{q},\alpha}^{\pm}
    &= \frac{1}{2\sqrt{N}}
    \sum_{\vv{k},\sigma,\sigma'}
    c_{\vv{k}+\vv{q},\alpha,\sigma}^{\dagger}
    \sigma^{\pm}_{\sigma\sigma'}
    c_{\vv{k},\alpha,\sigma'},
    \\
    c_{\vv{r}_{\alpha},\alpha\sigma}
    &= \frac{1}{\sqrt{N}}
    \sum_{\vv{k}}c_{\vv{k}\alpha\sigma}
    e^{-i\vv{k}\cdot\vv{r}_{\alpha}},
\end{align*}
where $N$ is the number of unit cells and $\sigma^{\pm}\equiv\sigma_x\pm i\sigma_y$.

To describe the type-II and type-III magnets on an equal footing, we
introduce two magnetic components, $A$ and $B$, in each unit cell. A
minimal model is
\begin{align}
    \mathcal{H}_{\rm AF}^{\rm spin}
    ={}&J_{0}\sum_{\vv{r}_A}
    \sum_{\vv{\delta}\in\mathcal{D}_{AB}}
    \vv{S}_{\vv{r}_A, A}\cdot
    \vv{S}_{\vv{r}_A+\vv{\delta}, B}
    \notag\\
    &- \frac{1}{2}\sum_{\vv{r},\mu=x,y}\sum_{\alpha=A,B}
    J_{\mu}^{\alpha}
    \vv{S}_{\vv{r}_{\alpha},\alpha}\cdot
    \vv{S}_{\vv{r}_{\alpha}+a\hat{\vv{\mu}},\alpha}
    \notag\\
    &-K\sum_{\vv{r},\alpha}
    \left(S_{\vv{r}_{\alpha},\alpha}^{z}\right)^{2},
    \qquad J_{0}>0 .
    \label{eq:spin_AF_general}
\end{align}
For sufficiently large $J_0$, its collinear ground state is
$\vv{S}_{\vv{r}_A, A}=S\hat{\vv{z}}$ and
$\vv{S}_{\vv{r}_B, B}=-S\hat{\vv{z}}$, and for all parameter sets used below we have numerically verified $\omega_{\vv{q}}^{\pm}>0$ throughout the Brillouin zone.
The two magnetic components are
equivalent in a type-II magnet, and hence
\begin{align}
    \mathcal{H}_{\rm II}^{\rm spin}
    =\left.\mathcal{H}_{\rm AF}^{\rm spin}\right|_{
    J_x^A=J_y^A=J_x^B=J_y^B=J}.
    \label{eq:spin_type_II}
\end{align}
Consequently, the two oppositely polarized magnon sectors have identical
dispersions, in analogy with the spin-degenerate bands of $H_{\rm II}$.

In the type-III case, the two components are not related by a translation
or inversion but are exchanged by a fourfold rotation. 
This is implemented by the altermagnetic exchange pattern
\begin{align}
    &J_x^A=J_y^B=J+\delta J,\qquad
    J_y^A=J_x^B=J-\delta J,
    \label{eq:spin_type_III_exchange}\\
    &\mathcal{H}_{\rm III}^{\rm spin}
    =\left.\mathcal{H}_{\rm AF}^{\rm spin}\right|_{
    \text{Eq.~\eqref{eq:spin_type_III_exchange}}}.
    \label{eq:spin_type_III}
\end{align}
Indeed, $C_{4z}$ interchanges $x$ and $y$ as well as $A$ and $B$, leaving
Eq.~\eqref{eq:spin_type_III} invariant.
The $\delta J \to 0$ limit recovers the type-II model with degenerate magnon branches, whereas finite $\delta J$ produces a $d$-wave splitting of the two oppositely polarized branches.
As shown in Appendix~\ref{app:AF_magnon}, and omitting an additive constant, the magnon Hamiltonian for the type-II and type-III magnets is written as
\begin{align}
    \mathcal{H}_{\rm AF}^{\rm mag}
    &=\sum_{\vv{q}}
    \left(\omega_{\vv{q}}^{+}\hat{\alpha}_{\vv{q}}^\dagger \hat{\alpha}_{\vv{q}}
    +\omega_{\vv{q}}^{-}\hat{\beta}_{\vv{q}}^\dagger \hat{\beta}_{\vv{q}}\right),
    \\
    \omega_{\vv{q}}^{\pm}
    &= E_{\vv{q}} \mp S\delta J \lb \cos(a q_x) - \cos(a q_y) \rb,
    \\
    \gamma_{0,\vv{q}}^{\rm AF}
    &= J_{0}S\Gamma_{\vv{q}},
    \\
    E_{\vv{q}} &= \sqrt{\gamma_{\vv{q}}^2 - (\gamma_{0,\vv{q}}^{\rm AF})^2},
    \quad
    u_{\vv{q}} = \sqrt{\frac{\gamma_{\vv{q}} + E_{\vv{q}}}{2E_{\vv{q}}}},
    \notag\\
    v_{\vv{q}} &= -\sgn(\gamma_{0,\vv{q}}^{\rm AF})\sqrt{\frac{\gamma_{\vv{q}} - E_{\vv{q}}}{2E_{\vv{q}}}}.
\end{align}
The energy spectra of antiferromagnetic and altermagnetic magnons are shown in Fig.~\ref{fig:energy}(e) and (f), respectively, where the two magnon modes are degenerate at the $\Gamma$ point in the presence of the magnetic anisotropy $K$.
The altermagnetic magnon has the $d$-wave splitting of the two magnon modes, which is similar to the electronic band structure of the type-III magnet.

Next, we derive the magnon-electron interaction Hamiltonian for the type-II and type-III magnets by using the Holstein-Primakoff transformation of Eq.~\eqref{eq:exchange_interaction}.
Retaining terms through first order in the magnon operators and neglecting the subleading longitudinal $b^\dagger b s^z$ terms, we obtain
\begin{align}
    \mathcal{H}_{\rm ex}
    &\simeq-J_{\rm ex}S\sum_{\vv{r}} \lb s_{\vv{r}_A,A}^{z} - s_{\vv{r}_B,B}^{z} \rb
    \notag\\
    &-\frac{J_{\rm ex}\sqrt{2S}}{2}\sum_{\vv{r}}
    \left[ b_{\vv{r}_A,A}^\dagger s_{\vv{r}_A,A}^{+}
    + b_{\vv{r}_A,A} s_{\vv{r}_A,A}^{-}
    \right.
    \notag\\
    &\qquad\qquad
    \left.
    + b_{\vv{r}_B,B} s_{\vv{r}_B,B}^{+}
    + b_{\vv{r}_B,B}^\dagger s_{\vv{r}_B,B}^{-} \right].
\end{align}
The first term corresponds to the exchange field $M\sigma_z\otimes\tau_z$ in $H_{\rm II}$ and $H_{\rm III}$, and we set $M = -J_{\rm ex}S/2$.
Thus, the magnon creating and annihilating interaction in the Fourier space is written as
\begin{align}
    \mathcal{H}_{\rm AF,int}
    ={}& \sqrt{\frac{2}{S}} M \sum_{\vv{q}}
    \bigl\{ b_{\vv{q}A}^\dagger s_{-\vv{q},A}^{+}
    + b_{\vv{q}A} s_{\vv{q},A}^{-}
    \notag\\
    &\qquad
    + b_{-\vv{q}B} s_{-\vv{q},B}^{+}
    + b_{-\vv{q}B}^\dagger s_{\vv{q},B}^{-} \bigr\},
    \notag\\
    ={}& \sqrt{\frac{2}{S}} M \sum_{\vv{q}}
    \left\{
    \lb u_{\vv{q}}s_{-\vv{q},A}^{+}
    + v_{\vv{q}}s_{-\vv{q},B}^{+} \rb \hat{\alpha}_{\vv{q}}^\dagger
    \right.
    \notag\\
    &\qquad
    + \lb u_{\vv{q}}s_{\vv{q},A}^{-}
    + v_{\vv{q}}s_{\vv{q},B}^{-} \rb \hat{\alpha}_{\vv{q}}
    \notag\\
    &\qquad
    + \lb v_{\vv{q}}s_{-\vv{q},A}^{+}
    + u_{\vv{q}}s_{-\vv{q},B}^{+} \rb \hat{\beta}_{\vv{q}}
    \notag\\
    &\qquad\left.
    + \lb v_{\vv{q}}s_{\vv{q},A}^{-}
    + u_{\vv{q}}s_{\vv{q},B}^{-} \rb \hat{\beta}_{\vv{q}}^\dagger
    \right\}.
\end{align}

Next, we define the electron and magnon propagators in the Matsubara formalism.
First, we introduce the electron propagator as
\begin{align}
    G_{\alpha\beta,\sigma\sigma'}^{(0)}(\vv{k},i\epsilon_n;\mathcal N)
    &= \delta_{\alpha\beta}\delta_{\sigma\sigma'}
    \frac{1}{i\epsilon_n - \xi_{\vv{k}\alpha\sigma}(\mathcal N)},
\end{align}
where $\xi_{\vv{k}\alpha\sigma}(\mathcal N) = \epsilon_{\vv{k}\alpha\sigma}(\mathcal N) - \mu$ is the electron energy measured from the chemical potential $\mu$.

Second, we introduce the magnon propagator.
Let $\mathcal{N}=+1$ denote the magnetic domain used in the
Holstein--Primakoff transformations above and $\mathcal{N}=-1$ its
time-reversed domain.  
We use the convention $D=-\langle T_\tau B(\tau)B^\dagger(0)\rangle$, where $\zeta_{m,\mathcal N}=+1$ and $-1$ select the annihilation and creation components, respectively.
Throughout, $\omega_{m,\vv{q}}>0$ denotes the energy of the magnon branch $m$, that is, $\gamma_{\vv{q}}+m\gamma_{0,\vv{q}}^{\rm I}$ for the type-I magnet and $\omega_{\vv{q}}^{m}$ for the type-II and type-III magnets.
We introduce the signed magnon propagator
\begin{align}
    D_{m,\mathcal{N}}(\vv{q},i\nu_{\ell})
    &=
    \frac{\zeta_{m,\mathcal{N}}}
    {i\nu_{\ell}-\zeta_{m,\mathcal{N}}\omega_{m,\vv{q}}},
    \\
    \zeta_{m,\mathcal{N}}
    &=
    \begin{cases}
        \mathcal{N},
        & \text{type I},\quad m=\pm,
        \\
        \mathcal{N},
        & \text{type II and type III},\quad m=+\;(\hat{\alpha}),
        \\
        -\mathcal{N},
        & \text{type II and type III},\quad m=-\;(\hat{\beta}).
    \end{cases}
\end{align}
The identifications $m=+\leftrightarrow\hat{\alpha}$ and $m=-\leftrightarrow\hat{\beta}$ refer to the $\mathcal{N}=+1$ domain; for $\mathcal{N}=-1$, the same $m$ labels the corresponding time-reversed branch with the same positive energy.
Thus, time reversal changes the pole orientation as
$\zeta_{m,-\mathcal{N}}=-\zeta_{m,\mathcal{N}}$, even when the positive
magnon energy $\omega_{m,\vv{q}}$ is unchanged.

\section{Charge-spin conversion matrix}\label{sec:matrix}
Here, we derive the charge-spin conversion matrix from the spin-resolved conductivities as
\begin{align}
    \mqty( 
        J^{\rm e}_{\mu}
        \\
        J^{\rm s}_{\mu}
     )
     =
        \mqty(
            \sigma_{\mu\nu}^{\rm e-e} & \sigma_{\mu\nu}^{\rm e-s}
            \\
            \sigma_{\mu\nu}^{\rm s-e} & \sigma_{\mu\nu}^{\rm s-s}
        )
        \mqty(
            E^{\rm e}_{\nu}
            \\
            E^{\rm s}_{\nu}
        ),
\end{align}
Here, $J^{\rm e}_{\mu} = J^{\rm e}_{\mu\up} + J^{\rm e}_{\mu\down}$ is the total charge current, whereas $J^{\rm s}_{\mu} = J^{\rm e}_{\mu\up} - J^{\rm e}_{\mu\down}$ is the spin-polarized charge current.
For $e>0$, the corresponding angular-momentum current is $J_{\mu,{\rm spin}}=-\hbar J^{\rm s}_{\mu}/(2e)$.
The conjugate fields are the common electric field $E^{\rm e}_{\nu} = (E^{\rm e}_{\nu\up} + E^{\rm e}_{\nu\down})/2$ and the spin electric field $E^{\rm s}_{\nu} = (E^{\rm e}_{\nu\up} - E^{\rm e}_{\nu\down})/2$.
The spin electric field $E^{\rm s}_{\nu}$ is understood as the gradient of a spin-dependent electrochemical potential generated by a nonequilibrium spin accumulation,
$E^{\rm s}_{\nu} = -\del_{\nu}\mu_{\rm s}/e$ with $\mu_{\rm s} = (\mu_{\up}-\mu_{\down})/2$.
Because the electron-magnon interaction of Eq.~\eqref{eq:exchange_interaction} transfers spin between the electron and the localized-spin subsystems, the spin-resolved densities are not separately conserved and obey
$\del_{t} n_{\vv{r}\sigma} + \boldsymbol{\nabla}\cdot\vv{j}_{\sigma} = \sigma \mathcal{T}$
with the electron-magnon spin-transfer torque $\mathcal{T}$, while the total charge density $n_{\up}+n_{\down}$ is conserved.
The spin-dependent electrochemical potential is therefore well defined only on time and length scales shorter than the spin relaxation set by $\mathcal{T}$, and the strict dc limit of $\sigma^{\rm s-s}_{\mu\nu}$ must be understood with this restriction.
For this reason we present the response matrix at finite frequency throughout this paper, where the restriction is satisfied as long as $\omega$ exceeds the spin relaxation rate \cite{Cite-in-Almag-S-E-conv_2006_Shi}.
The conductivities are defined as 
\begin{align}
    \sigma_{\mu\nu}^{\rm e-e} &= \sigma_{\mu\nu,\up\up} + \sigma_{\mu\nu,\up\down} + \sigma_{\mu\nu,\down\up} + \sigma_{\mu\nu,\down\down},
    \\  
    \sigma_{\mu\nu}^{\rm e-s} &= \sigma_{\mu\nu,\up\up} - \sigma_{\mu\nu,\up\down} + \sigma_{\mu\nu,\down\up} - \sigma_{\mu\nu,\down\down},
    \\
    \sigma_{\mu\nu}^{\rm s-e} &= \sigma_{\mu\nu,\up\up} + \sigma_{\mu\nu,\up\down} - \sigma_{\mu\nu,\down\up} - \sigma_{\mu\nu,\down\down},
    \\
    \sigma_{\mu\nu}^{\rm s-s} &= \sigma_{\mu\nu,\up\up} - \sigma_{\mu\nu,\up\down} - \sigma_{\mu\nu,\down\up} + \sigma_{\mu\nu,\down\down},
\end{align}
where $\sigma_{\mu\nu,\sigma\sigma'}$ is the conductivity for the electron current with spin $\sigma$ under the electric field with spin $\sigma'$.
For $\omega>0$, the regular dissipative part of the conductivity tensor is calculated from the Kubo formula as
\begin{align}
    \mathrm{Re}\sigma_{\mu\nu,\sigma\sigma'}^{\rm reg}(\omega)
    = \frac{\mathrm{Im}\Phi_{\mu\nu,\sigma\sigma'}^{\rm R}(\omega)}{\omega},
\end{align}
The zero-frequency Drude/contact contribution proportional to $\delta(\omega)$ is not included.
Here we adopt the convention $\Phi_{\mu\nu,\sigma\sigma'}^{\rm R}(t)=+i\theta(t)\langle[J_{\mu\sigma}(t),J_{\nu\sigma'}(0)]\rangle$, corresponding to the positive Matsubara correlation function defined below.
In the Matsubara formalism, the current-current correlation function is 
\begin{align}
    \Phi_{\mu\nu,\sigma\sigma'}(i\Omega_\lambda)
    = \int_0^{\beta_{T}} d\tau e^{i\Omega_\lambda\tau}
    \langle T_\tau J_{\mu\sigma}(\tau) J_{\nu\sigma'}(0) \rangle,
\end{align}
where $J_{\mu\sigma}(\tau) = e^{\tau \mathcal{H}} J_{\mu\sigma} e^{-\tau \mathcal{H}}$ is the current operator in the Heisenberg picture, and $\Omega_\lambda = 2\pi\lambda/\beta_{T}$ is the bosonic Matsubara frequency with $\beta_{T} = 1/k_{B}T$.

\begin{figure}
    \centering 
    \subfigure[Spin diagonal correlation]{
        \includegraphics[scale=1.0]{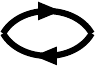}
        \label{fig:one-loop}
    }
    \subfigure[Spin flip correlation]{
        \includegraphics[scale=0.7]{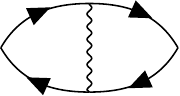}
        \label{fig:one-lang}
    }
    \caption{
        Feynman diagrams for the current-current correlation function.
        (a) The one-loop diagram for the spin-diagonal correlation function.
        (b) The one-rung diagram for the spin-flip correlation function.
    }
    \label{fig:diagram}
\end{figure}

We show the Feynman diagrams for the current-current correlation function in Fig.~\ref{fig:diagram}.
Figure~\ref{fig:one-loop} shows the one-loop diagram for the spin-diagonal correlation function, and Fig.~\ref{fig:one-lang} shows the one-rung diagram for the spin-flip correlation function.
In these diagrams, the thick solid line represents the dressed Matsubara electron propagator $G_{\alpha,\sigma}(\vv{k},i\epsilon_n;\mathcal{N})$, the thin solid line represents the bare electron propagator $G_{\alpha,\sigma}^{(0)}(\vv{k},i\epsilon_n;\mathcal{N})$, and the wavy line represents the magnon propagator $D_{m,\mathcal{N}}(\vv{q},i\nu_\ell)$.

Within the self-energy-dressed bubble approximation, with the electron self-energy derived in Appendix~\ref{app:self_energy}, the spin-diagonal correlation function is given by
\begin{align}
    \Phi_{\mu\nu,\sigma\sigma}(i\Omega_{\lambda}; \mathcal{N})
    &=
    -\frac{1}{\beta_{T}N}\sum_{\vv{k},\epsilon_n,\alpha}
    j_{\mu,\alpha\sigma}(\vv{k})
    j_{\nu,\alpha\sigma}(\vv{k})
    \notag\\
    &\times
    G_{\alpha,\sigma}(\vv{k},i\epsilon_n; \mathcal{N})
    G_{\alpha,\sigma}(\vv{k},i\epsilon_n+i\Omega_{\lambda}; \mathcal{N}),
\end{align}
where $j_{\mu,\alpha\sigma}(\vv{k}) = -e\del_{k_\mu}\xi_{\vv{k}\alpha\sigma}$ is the current vertex.
As shown in Appendix~\ref{app:spin_diagonal_conductivity}, the spin-diagonal conductivity is given by
\begin{align}
    \mathrm{Re} \sigma_{\mu\nu,\sigma\sigma}(\omega; \mathcal{N})
    &=
    \frac{\pi}{\omega N}\sum_{\vv{k},\alpha}
    j_{\mu,\alpha\sigma}(\vv{k})
    j_{\nu,\alpha\sigma}(\vv{k})
    \notag\\    
    &\times
    \int d\epsilon
    A_{\alpha,\sigma}(\vv{k},\epsilon; \mathcal{N})
    A_{\alpha,\sigma}(\vv{k},\epsilon+\omega; \mathcal{N})
    \notag\\
    &\times
    \left[
        f(\epsilon) - f(\epsilon+\omega)
    \right],
\end{align} 
where $A_{\alpha,\sigma}(\vv{k},\epsilon; \mathcal{N}) = -\frac{1}{\pi}\mathrm{Im} G_{\alpha,\sigma}^{\rm R}(\vv{k},\epsilon; \mathcal{N})$ is the electron spectral function normalized as $\int d\epsilon\, A_{\alpha,\sigma}(\vv{k},\epsilon; \mathcal{N}) = 1$, and $f(\epsilon) = 1/(e^{\beta_{T}\epsilon}+1)$ is the Fermi distribution function.

Next, we derive the spin-flip conductivity $\sigma_{\mu\nu,\sigma\sigma'}$ by using the one-rung diagram.
The one-rung diagram is written as
\begin{align}
    &\Phi_{\mu\nu,\sigma\sigma'}(i\Omega_{\lambda}; \mathcal{N})
    \notag\\
    &=
    \frac{1}{\beta_{T}^2N^2}\sum_{\vv{k},\vv{q},\epsilon_n,\nu_\ell,\alpha,m}
    j_{\mu,\alpha\sigma}(\vv{k})
    j_{\nu,\alpha\sigma'}(\vv{k}+\sigma\vv{q})
    \notag\\
    &\times
    C_{\alpha, m}(\vv{q})G_{\alpha,\sigma}^{(0)}(\vv{k},i\epsilon_n;\mathcal N)
    G_{\alpha,\sigma}^{(0)}(\vv{k},i\epsilon_n+i\Omega_{\lambda};\mathcal N)
    \notag\\
    &\times
    G_{\alpha,\sigma'}^{(0)}(\vv{k}+\sigma\vv{q},i\epsilon_n+i\sigma\nu_\ell;\mathcal N)
    D_{m,\mathcal{N}}(\vv{q},i\nu_\ell)
    \notag\\ 
    &\times
    G_{\alpha,\sigma'}^{(0)}(\vv{k}+\sigma\vv{q},i\epsilon_n+i\sigma\nu_\ell+i\Omega_{\lambda};\mathcal N),
\end{align}
where $\sigma' = -\sigma$.
$G_{\alpha,\sigma}^{(0)}(\vv{k},i\epsilon_n;\mathcal N)$ is the bare electron propagator, and $D_{m,\mathcal{N}}(\vv{q},i\nu_\ell)$ is the magnon propagator.
The overall sign differs from that of the spin-diagonal bubble because the single fermion loop and the convention $D=-\Lb T_{\tau}B B^{\dagger}\Rb$ each contribute a factor $-1$.
Here the electron propagators are the bare ones, whereas the spin-diagonal bubble is dressed by the magnon self-energy.
The two channels enter the observable $\sigma^{\rm e-e}_{\mu\nu}-\sigma^{\rm s-s}_{\mu\nu}=2\lb\sigma_{\mu\nu,\up\down}+\sigma_{\mu\nu,\down\up}\rb$ separately, in which the spin-diagonal contributions cancel identically, so that this asymmetry does not affect the spin-flip results discussed below.
The electron--magnon vertex weight $C_{\alpha, m}(\vv{q})=|g_{\alpha,m}(\vv{q})|^2$ is given by $C_{\alpha, m}(\vv{q}) = \Delta^2/S$ for the type-I magnet, and for the type-II and type-III magnets, it is given by $C_{\alpha, m}(\vv{q}) = |u_{\vv{q}}|^2 (2M^2/S)$ for $(m,\alpha) = (+,A), (-,B)$ and $C_{\alpha, m}(\vv{q}) = |v_{\vv{q}}|^2 (2M^2/S)$ for $(m,\alpha) = (-,A), (+,B)$.
As shown in Appendix~\ref{app:spin_flip_conductivity}, this correlation function can be calculated as 
\begin{align}
    \Phi_{\mu\nu,\sigma\sigma'}(i\Omega_{\lambda}; \mathcal{N})
    &=
    \frac{1}{N^2}\sum_{\vv{k},\vv{q},\alpha,m}
    j_{\mu,\alpha\sigma}(\vv{k})
    j_{\nu,\alpha\sigma'}(\vv{k}+\sigma\vv{q})
    \notag\\
    \times&
    C_{\alpha, m}(\vv{q})
    \mathcal{K}_{\sigma,m,\mathcal{N}}(\xi_{\vv{k}\alpha\sigma},\xi_{\vv{k}+\sigma\vv{q},\alpha\sigma'};i\Omega_\lambda),
\end{align}
where the kernel function is given as 
\begin{align}
    \mathcal{K}_{\sigma,m,\mathcal{N}}(a,b;z)
    &=
    \frac{2\sigma \zeta_{m,\mathcal{N}}(f(b) - f(a))}{(\sigma\Delta E - \zeta_{m, \mathcal{N}}\omega_{m,\vv{q}})^2 - z^2}
    \notag\\ 
    &\times
    \frac{n(\zeta_{m,\mathcal{N}}\omega_{m,\vv{q}}) - n(\sigma \Delta E)}{\sigma \Delta E - \zeta_{m, \mathcal{N}}\omega_{m,\vv{q}}},
\end{align}
where $\Delta E = b - a$ and $n(\epsilon) = 1/(e^{\beta_{T}\epsilon}-1)$ is the Bose distribution function.
According to the analytic continuation $i\Omega_{\lambda} \to \omega + i0$ and the Kubo formula, the spin-flip conductivity is given by
\begin{align}
    \mathrm{Re}\,\sigma_{\mu\nu,\sigma\sigma'}(\omega; \mathcal{N})
    &=
    \frac{\mathrm{Im}\Phi_{\mu\nu,\sigma\sigma'}^{\rm R}(\omega; \mathcal{N})}{\omega},
\end{align}
where $\Phi_{\mu\nu,\sigma\sigma'}^{\rm R}(\omega; \mathcal{N})$ is the retarded correlation function obtained by the analytic continuation $i\Omega_{\lambda} \to \omega + i0$.

\subsection{Magnetic-domain parity of the response matrix}\label{sec:parity}
Reversing the magnetic domain, $\mathcal{N}\to-\mathcal{N}$, flips the sign of the exchange field, $\Delta \to -\Delta$ and $M \to -M$, so that the two spin sectors of the electron spectrum are interchanged,
\begin{align}
    \xi_{\vv{k}\alpha\sigma}(-\mathcal{N}) = \xi_{\vv{k}\alpha\bar{\sigma}}(\mathcal{N}),
    \qquad \bar{\sigma} = -\sigma .
\end{align}
The magnon energies $\omega_{m,\vv{q}}$ and the interaction vertices $C_{\alpha,m}(\vv{q})$ are even in $\Delta$ and $M$ and hence unchanged, whereas the pole orientation of the magnon propagator reverses, $\zeta_{m,-\mathcal{N}} = -\zeta_{m,\mathcal{N}}$.
The self-energy and spectral function transform accordingly as $\Sigma_{\alpha,\sigma}^{\rm R}(\vv{k},\epsilon; -\mathcal{N}) = \Sigma_{\alpha,\bar{\sigma}}^{\rm R}(\vv{k},\epsilon; \mathcal{N})$ and $A_{\alpha,\sigma}(\vv{k},\epsilon; -\mathcal{N}) = A_{\alpha,\bar{\sigma}}(\vv{k},\epsilon; \mathcal{N})$.
Applying these relations to the spin-diagonal bubble and the kernel $\mathcal{K}_{\sigma,m,\mathcal{N}}$, together with the transformation of the current vertices, we find that the domain reversal acts as a plain exchange of the spin labels,
\begin{align}
    \sigma_{\mu\nu,\sigma\sigma}(\omega; -\mathcal{N})
    &= \sigma_{\mu\nu,\bar{\sigma}\bar{\sigma}}(\omega; \mathcal{N}),
    \\
    \sigma_{\mu\nu,\sigma\bar{\sigma}}(\omega; -\mathcal{N})
    &= \sigma_{\mu\nu,\bar{\sigma}\sigma}(\omega; \mathcal{N}).
\end{align}
Substituting these relations into the definitions of the conversion matrix gives the domain parity summarized in Table~\ref{tab:parity}: the diagonal blocks $\sigma^{\rm e-e}$ and $\sigma^{\rm s-s}$ are even in $\mathcal{N}$, while the off-diagonal blocks $\sigma^{\rm e-s}$ and $\sigma^{\rm s-e}$ are odd.

\begin{table}[tb]
    \caption{
        Parity of the charge-spin conversion matrix under the magnetic-domain reversal $\mathcal{N}\to-\mathcal{N}$.
        The last column gives the combination of the spin-resolved conductivities that survives for the diagonal components $\mu=\nu$, where $\Phi_{\mu\mu,\up\down} = \Phi_{\mu\mu,\down\up}$ holds.
    }
    \label{tab:parity}
    \begin{ruledtabular}
    \begin{tabular}{lcc}
        Response & $\mathcal{N}$ parity & diagonal component \\
        \hline
        $\sigma^{\rm e-e}_{\mu\mu}$ & even & $\sigma_{\up\up}+\sigma_{\down\down}+2\sigma_{\up\down}$ \\
        $\sigma^{\rm s-s}_{\mu\mu}$ & even & $\sigma_{\up\up}+\sigma_{\down\down}-2\sigma_{\up\down}$ \\
        $\sigma^{\rm e-s}_{\mu\mu}$ & odd  & $\sigma_{\up\up}-\sigma_{\down\down}$ \\
        $\sigma^{\rm s-e}_{\mu\mu}$ & odd  & $\sigma_{\up\up}-\sigma_{\down\down}$ \\
    \end{tabular}
    \end{ruledtabular}
\end{table}

Two consequences follow.
First, together with $\Phi_{\mu\mu,\up\down} = \Phi_{\mu\mu,\down\up}$, the diagonal components satisfy $\sigma^{\rm e-s}_{\mu\mu} = \sigma^{\rm s-e}_{\mu\mu}$.
Combined with their odd domain parity, this yields the diagonal Onsager--Casimir relation $\sigma^{\rm e-s}_{\mu\mu}(\omega;\mathcal{N}) = -\sigma^{\rm s-e}_{\mu\mu}(\omega;-\mathcal{N})$.
Second, the magnon-mediated spin-flip combination
$\sigma^{\rm e-e}_{\mu\nu}-\sigma^{\rm s-s}_{\mu\nu} = 2(\sigma_{\mu\nu,\up\down}+\sigma_{\mu\nu,\down\up})$,
and hence the $B_{1g}$ strain response $\chi^{B_{1g}}_{\rm sf}$ introduced in Sec.~\ref{sec:b1g}, are even in $\mathcal{N}$, whereas the charge-spin conversion $\sigma^{\rm e-s}_{\mu\nu}$ is odd.
The coupling modulation introduced in Sec.~\ref{sec:b1g} only rescales the spin-independent, $\vv{q}$-even vertex weights, so $\sigma_{\mu\mu,\up\down}^{(\epsilon)} = \sigma_{\mu\mu,\down\up}^{(\epsilon)}$ remains valid at every $\epsilon$; consequently, $\chi_{\rm sf}^{B_{1g}}$ is also even under $\mathcal{N}\to-\mathcal{N}$.
In a sample with equal populations of the inverse magnetic domains $\mathcal{N}=\pm1$, the charge-spin conversion averages out, whereas the magnon-mediated spin-flip channel survives.
Thus, this channel does not require selecting one of the inverse magnetic domains.
This statement does not apply to rotational twins related by a $\pi/2$ rotation, whose averaging can cancel the $B_{1g}$ response.

\begin{figure}
    \centering 
    \includegraphics[scale=1.0]{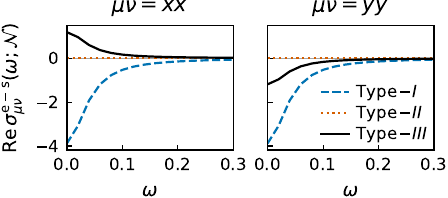}
    \caption{
        The real part of the regular charge-spin conversion conductivity, $\mathrm{Re}\,\sigma^{\rm e-s}_{\mu\nu}(\omega; \mathcal{N})$, in units of $e^2/\hbar$ for the type-I, type-II, and type-III magnets.
        In this figure, we set the magnetic domain $\mathcal{N} = +1$, temperature $T = 0.2$, the chemical potential $\mu = 0.5$, and the other parameters are set as $t = J = S = a = 1$, $t' = 0.2$, $\Delta = M = 0.3$, $J_{0} = 0.4$, $K=0.1$ and $\delta J = 0.2$.
        The left and right panels show the $\mu=\nu=x$ and $\mu=\nu=y$ components of it, respectively.
        The leftmost data point is evaluated at $\omega_{\min}=0.001$, the smallest positive frequency used in the calculation.
    }
    \label{fig:charge_spin_e_s}
\end{figure}

\section{Results}\label{sec:results}
In this section, we show the numerical results for the conductivities.
Throughout this section, $\sigma$ and all response coefficients derived from it denote the real part of the regular conductivity at $\omega>0$, unless otherwise stated.
The spin-diagonal and spin-flip responses were evaluated with different numerical quadratures; implementation details and convergence tests are given in Appendix~\ref{app:numerics}.

\subsection{Anisotropy of the charge-spin conversion}
With our matrix convention, $\sigma_{\mu\mu}^{\rm e-s}$ describes the inverse response from a spin electric field to a charge current, whereas $\sigma_{\mu\mu}^{\rm s-e}$ describes the direct charge-to-spin response.
Since these longitudinal coefficients are equal, we use $\sigma_{\mu\mu}^{\rm e-s}$ to represent both responses.
Figure~\ref{fig:charge_spin_e_s} shows the frequency dependence of the charge-spin conversion matrix $\sigma^{\rm e-s}_{\mu\nu}(\omega; \mathcal{N})$ for the type-I, type-II, and type-III magnets.
From this figure, we can see that the type-I has the isotropic charge-spin conversion, and the type-III has the anisotropic charge-spin conversion, while the type-II has no charge-spin conversion.
For the diagonal components, the change of variables $\vv{k}'=\vv{k}+\sigma\vv{q}$ exchanges the two electron energies and current vertices.
Together with $\mathcal{K}_{-\sigma,m,\mathcal{N}}(b,a;z)=\mathcal{K}_{\sigma,m,\mathcal{N}}(a,b;z)$, this gives $\Phi_{\mu\mu,\uparrow\downarrow}^{\rm R}(\omega;\mathcal{N})=\Phi_{\mu\mu,\downarrow\uparrow}^{\rm R}(\omega;\mathcal{N})$.
Thus, the longitudinal charge-spin conversion is written by the spin-diagonal conductivities only as
\begin{align}
    \sigma^{\rm e-s}_{\mu\mu}(\omega; \mathcal{N})
    &= \sigma_{\mu\mu,\uparrow\uparrow}(\omega; \mathcal{N}) - \sigma_{\mu\mu,\downarrow\downarrow}(\omega; \mathcal{N}).
\end{align}
Also, for the type-II magnet, the two spin sectors are equivalent, and thus $\sigma^{\rm e-s}_{\mu\mu}(\omega; \mathcal{N}) = 0$.

\begin{figure}
    \centering 
    \includegraphics[scale=1.0]{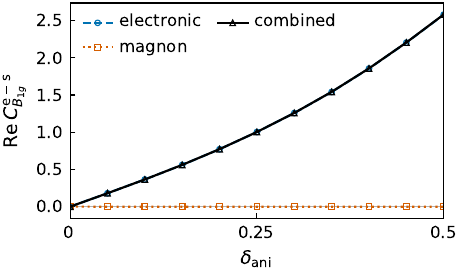}
    \caption{
        The real part of the $B_{1g}$ anisotropy of the charge-spin conversion, $\mathrm{Re}\,C_{B_{1g}}^{\rm e-s}$, for the type-III magnet at $\omega=0.03$, in units of $e^2/\hbar$.
        The electronic, magnon, and combined curves correspond to $(t',\delta J)=(\delta_{\rm ani},0)$, $(0,\delta_{\rm ani})$, and $(\delta_{\rm ani},\delta_{\rm ani})$, respectively.
        We set $\mathcal{N}=+1$, $T=0.2$, $\mu=0.5$, $t=J=S=a=1$, $\Delta=M=0.3$, $J_0=0.4$, and $K=0.1$.
    }
    \label{fig:anisotropy_es_b1g_attribution}
\end{figure}

Next, we discuss the origin of the anisotropy of the charge-spin conversion matrix $\sigma^{\rm e-s}_{\mu\nu}(\omega; \mathcal{N})$ for the type-III magnet.
We define $C^{\rm e-s}_{B_{1g}} = \sigma^{\rm e-s}_{xx} - \sigma^{\rm e-s}_{yy}$ as the anisotropy of the charge-spin conversion matrix.
It vanishes in the isotropic limit $\delta_{\rm ani}=0$, where the type-III model reduces to the type-II one.
To compare the electronic and magnonic sources on an equal footing, we vary their relative anisotropies $t'/t$ and $\delta J/J$ by the common parameter $\delta_{\rm ani}$.
As the anisotropy of the model, we consider the cases $t' = \delta_{\rm ani}$ and $\delta J = 0$, $t' = 0$ and $\delta J = \delta_{\rm ani}$, and $t' = \delta J = \delta_{\rm ani}$.
The first case corresponds to the anisotropy of the electron system, the second case corresponds to the anisotropy of the magnon system, and the third case corresponds to the anisotropy of both systems.
Figure~\ref{fig:anisotropy_es_b1g_attribution} shows the anisotropy of the charge-spin conversion matrix $C^{\rm e-s}_{B_{1g}}$ for the type-III magnet.
From this figure, we can see that the anisotropy of the charge-spin conversion matrix is mainly attributed to the anisotropy of the electron system, and the anisotropy of the magnon system has a minor contribution to it.
Thus, we can conclude that the anisotropy of the charge-spin conversion matrix originates in the spin splitting of the altermagnetic electron system, the same nonrelativistic splitting that underlies the spin-splitter effect.
This attribution holds within the present leading-order treatment, in which the magnons enter the spin-diagonal channel only through the electron self-energy; vertex corrections of the same order are not included.

\subsection{Spin flip conductivity mediated by magnons}\label{sec:spinflip}

\begin{figure}
    \centering 
        \includegraphics[scale=0.9]{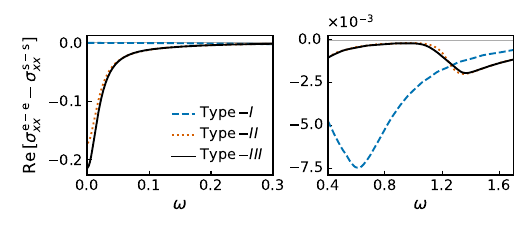}
    \caption{
        The frequency dependence of the difference between the charge--charge and spin--spin responses, $\mathrm{Re}[\sigma^{\rm e-e}_{xx}(\omega; \mathcal{N}) - \sigma^{\rm s-s}_{xx}(\omega; \mathcal{N})]$, in units of $e^2/\hbar$ for the type-I, type-II, and type-III magnets.
        In this figure, we set the magnetic domain $\mathcal{N} = +1$, temperature $T = 0.2$, the chemical potential $\mu = 0.5$, and the other parameters are set as $t = J = S = a = 1$, $t' = 0.2$, $\Delta = M = 0.3$, $J_{0} = 0.4$, $K=0.1$ and $\delta J = 0.2$.
        The left panel shows the low-frequency region $\omega = 0.001$ to $0.3$, where $\omega_{\min}=0.001$ is the smallest positive frequency used in the calculation.
        The right panel shows the high frequency region $\omega = 0.4$ to $1.7$.
    }
    \label{fig:ee_minus_ss_xx_vs_omega}
\end{figure}

Next, we discuss the difference between the charge--charge conductivity and the spin--spin conductivity
\begin{align}
    &\sigma^{\rm e-e}_{\mu\nu}(\omega;\mathcal{N})
    -\sigma^{\rm s-s}_{\mu\nu}(\omega;\mathcal{N})
    \notag\\
    &\qquad=
    2\left[
        \sigma_{\mu\nu,\uparrow\downarrow}(\omega;\mathcal{N})
        +\sigma_{\mu\nu,\downarrow\uparrow}(\omega;\mathcal{N})
    \right].
\end{align}
Figure~\ref{fig:ee_minus_ss_xx_vs_omega} shows the frequency dependence of the difference between the charge--charge conductivity and the spin--spin conductivity for the type-I, type-II, and type-III magnets.
For the parameters used here, the type-II and type-III magnets show closely similar frequency dependence in the real spin-flip response.
Writing $S_{{\rm sf},\mu\mu}=\sigma_{\mu\mu}^{\rm e-e}-\sigma_{\mu\mu}^{\rm s-s}$, a $\pi/2$ rotation gives $S_{{\rm sf},xx}(t',\delta J)=S_{{\rm sf},yy}(-t',-\delta J)$.
In the unperturbed type-III model, the spin-flip $C_4$ symmetry together with $\Phi_{\mu\mu,\uparrow\downarrow}^{\rm R}=\Phi_{\mu\mu,\downarrow\uparrow}^{\rm R}$ also gives $S_{{\rm sf},xx}=S_{{\rm sf},yy}$.
Hence $S_{{\rm sf},xx}(t',\delta J)=S_{{\rm sf},xx}(-t',-\delta J)$, so the linear correction along $t'/t=\delta J/J=\delta_{\rm ani}$ vanishes, although the quadratic terms $t'^2$, $t'\delta J$, and $\delta J^2$ remain.
The right panel of Fig.~\ref{fig:ee_minus_ss_xx_vs_omega} resolves the higher-frequency region, where the signed real response exhibits a negative dip, or a peak in magnitude, near $\omega\simeq0.6$ for the type-I magnet and near $\omega\simeq1.4$ for the type-II and type-III magnets.
These structures are not onset thresholds, because a finite response is already present at lower frequencies.
The resonance condition $\omega=\left|\sigma\Delta E-\zeta_{m,\mathcal{N}}\omega_{m,\vv{q}}\right|$ shows that they arise from composite electron--magnon excitations rather than from the magnon gap alone.
For $\mathcal{N}=+1$ and $\vv{q}\to0$, the spin-flip transition energies reduce to
\begin{align}
    \Delta E_{\rm I}&=-2\sigma\Delta,
    \\
    \Delta E_{\rm II/III}&=-2\sigma\eta_\alpha M.
\end{align}
For the low magnon branch of the type-I magnet and the $u_{\vv{q}}^2$ channel of the antiferromagnetic magnets, these give the sum-energy references
\begin{align}
    \omega_{\rm sum}^{\rm I}&=2|\Delta|+\omega_{\rm gap},
    \\
    \omega_{\rm sum}^{\rm AF}&=2|M|+\omega_{\rm gap},
\end{align}
whereas the $v_{\vv{q}}^2$ channel of the antiferromagnetic magnets gives the difference-energy reference
\begin{align}
    \omega_{\rm diff}^{\rm AF}
    =\left|\omega_{\rm gap}-2\lvert M\rvert\right|.
\end{align}
Here, $\omega_{\rm gap}$ denotes the relevant magnon energy at $\vv{q}=0$.
The $\vv{q}=0$ sum-energy references are $2|\Delta|+2KS=0.8$ for the type-I magnet and $2|M|+S\sqrt{2K(8J_{0}+2K)}\simeq1.42$ for the type-II and type-III magnets.
The latter agrees closely with the dip near $\omega\simeq1.4$, whereas the type-I value provides only a qualitative energy scale for the dip near $\omega\simeq0.6$.
The integrated feature position need not coincide with the $\vv{q}=0$ reference because finite-$\vv{q}$ transition energies, occupation factors, and vertex weights all contribute; the comparatively flat type-I magnon branch permits a larger shift.
This resonance structure makes the magnon energy scale visible in the frequency dependence.
The two panels of Fig.~\ref{fig:ee_minus_ss_xx_vs_omega} are drawn on different vertical scales; the full frequency dependence on a common scale is given in Appendix~\ref{app:full_frequency}.

\subsection{Response to $B_{1g}$ strain}\label{sec:b1g}
\begin{figure}
    \centering 
    \includegraphics{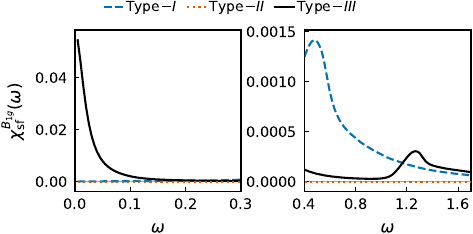}
    \caption{
        The frequency dependence of the $B_{1g}$ strain susceptibility of the spin-flip conductivity, $\chi^{B_{1g}}_{\rm sf}(\omega)$, in units of $e^{2}/\hbar$ for the type-I, type-II, and type-III magnets.
        The left panel shows the low-frequency region $0.001 \leq \omega \leq 0.3$, and the right panel the high-frequency region from $\omega=0.4$ to $1.7$; the two panels are drawn on different vertical scales.
        The susceptibility vanishes identically for the type-II magnet at every frequency, as shown in the text.
        Since $\sigma^{B_{1g}}_{\rm sf}=\chi^{B_{1g}}_{\rm sf}\epsilon$ holds exactly, $\chi^{B_{1g}}_{\rm sf}$ does not depend on $\epsilon$; the curves are evaluated at $\epsilon=1/3$.
        In this figure, we set the magnetic domain $\mathcal{N} = +1$, temperature $T = 0.2$, the chemical potential $\mu = 0.5$, and the other parameters are set as $t = J = S = a = 1$, $t' = 0.2$, $\Delta = M = 0.3$, $J_{0} = 0.4$, $K=0.1$ and $\delta J = 0.2$.
    }
    \label{fig:chi_b1g_frequency}
\end{figure}

Finally, to isolate the coupling contribution to the $B_{1g}$ strain response, we model the strain as a sublattice-selective modulation of the electron--magnon vertex weights while keeping the electron and magnon spectra unchanged.
A physical strain would also deform the electron and magnon dispersions, and those contributions are generally the larger ones.
Holding the spectra fixed therefore defines the scope of the present analysis: the symmetry of each subsystem is taken as given, and only the electron--magnon coupling is allowed to break the bulk $B_{1g}$ symmetry.
Evaluating the response that comes through the deformed spectra is left for future work.
Because $C_{4z}$ interchanges the two sublattices, a perturbation that lowers the bulk $B_{1g}$ symmetry necessarily appears as a sublattice-staggered, $\eta_{\alpha}$-odd modulation.
The interaction correction is represented by 
\begin{align}
    C^{(\epsilon)}_{A, m} = (1 + \epsilon) C_{A, m},\qquad
    C^{(\epsilon)}_{B, m} = (1 - \epsilon) C_{B, m},
    \label{eq:coupling_modulation}
\end{align}
where $\epsilon$ is the $B_{1g}$ strain parameter.
The one-rung response is linear in $C_{\alpha,m}^{(\epsilon)}$, so its dependence on $\epsilon$ is linear by construction; the distinguishing quantity is therefore its slope.
The equality $\sigma_{\mu\mu,\uparrow\downarrow}^{(\epsilon)}=\sigma_{\mu\mu,\downarrow\uparrow}^{(\epsilon)}$ means that the plotted single-channel quantity $\sigma_{\rm sf}^{B_{1g}}=\sigma_{xx,\downarrow\uparrow}-\sigma_{yy,\downarrow\uparrow}$ is one quarter of the $B_{1g}$ component of the full magnon-mediated spin-flip combination $\sigma^{\rm e-e}-\sigma^{\rm s-s}$.
Since the conductivity is linear in $C_{\alpha, m}^{(\epsilon)}$, we define the $B_{1g}$ strain susceptibility of the spin-flip conductivity $\chi^{B_{1g}}_{\rm sf}$ as
\begin{align}
    \sigma_{\rm sf}^{B_{1g}} = \chi^{B_{1g}}_{\rm sf} \epsilon.
\end{align}
This susceptibility is shown in Fig.~\ref{fig:chi_b1g_frequency}.
The left and right panels show the low and high frequency regions, respectively.
In the low frequency region, the susceptibility of the type-III magnet exceeds that of the type-I magnet by about two orders of magnitude, while that of the type-II magnet vanishes, as shown below.
In the high frequency region, the type-I and type-III magnets both develop a resonance structure.
It shares its origin with the dips of Fig.~\ref{fig:ee_minus_ss_xx_vs_omega}: the resonance condition $\omega=\left|\sigma\Delta E-\zeta_{m,\mathcal{N}}\omega_{m,\vv{q}}\right|$ places it at the combined energy scale of an electronic spin flip and a magnon excitation, not at the magnon gap alone.
As in that case, the feature lies somewhat below the $\vv{q}=0$ sum-energy references $\omega_{\rm sum}^{\rm I}$ and $\omega_{\rm sum}^{\rm AF}$, because finite-$\vv{q}$ transition energies, occupation factors, and vertex weights all contribute.
The ordering of the two magnets is reversed in this region, the type-I susceptibility exceeding the type-III one for $0.4\lesssim\omega\lesssim1.0$.
The separation of the type-III magnet from the type-I magnet by $\chi^{B_{1g}}_{\rm sf}$ is therefore restricted to the low frequency region, whereas the vanishing of $\chi^{B_{1g}}_{\rm sf}$ in the type-II magnet holds at every frequency.

\begin{figure}
    \centering 
    \renewcommand{\thesubfigure}{(\roman{subfigure})}
    \subfigure[$\omega = 0.03$]{
        \includegraphics{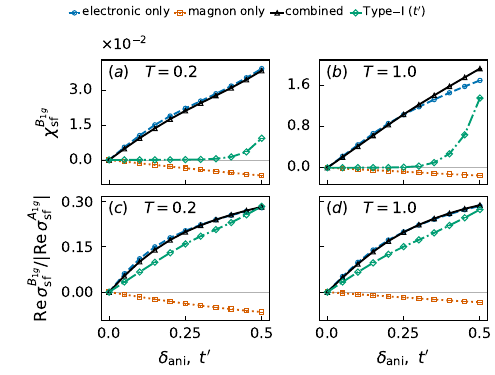}
        }
    \subfigure[$\omega = 1.3$]{
        \includegraphics{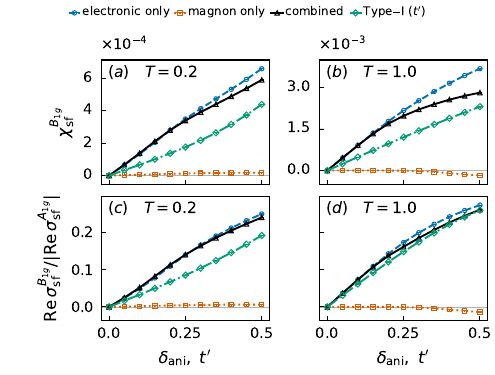}
    }
    \caption{
        Decomposition of the $B_{1g}$ strain susceptibility of the spin-flip conductivity into its electronic and magnonic sources, shown at $\omega=0.03$ in (i) and at $\omega=1.3$ in (ii).
        The parameter $\delta_{\rm ani}$ controls the relative anisotropies of the electron and magnon systems: the electronic, magnon, and combined curves correspond to $(t',\delta J)=(\delta_{\rm ani},0)$, $(0,\delta_{\rm ani})$, and $(\delta_{\rm ani},\delta_{\rm ani})$, respectively, the last one being the type-III magnet.
        The type-I magnet is shown for comparison; its magnon branches carry no $d$-wave splitting, so only $t'$ is varied there, and the abscissa accordingly denotes $\delta_{\rm ani}$ for the first three curves and $t'$ for the type-I one.
        Within each of (i) and (ii), the upper row shows $\chi^{B_{1g}}_{\rm sf}$ in units of $e^{2}/\hbar$, and the lower row shows the $B_{1g}$ component normalized by the $A_{1g}$ one, $\mathrm{Re}\,\sigma^{B_{1g}}_{\rm sf}/|\mathrm{Re}\,\sigma^{A_{1g}}_{\rm sf}|$ with $\sigma^{A_{1g}}_{\rm sf}=\sigma_{xx,\downarrow\uparrow}+\sigma_{yy,\downarrow\uparrow}$, while the left and right columns show the results at $T=0.2$ and $T=1.0$, respectively.
        In this figure, we set the magnetic domain $\mathcal{N} = +1$, the chemical potential $\mu = 0.5$, and the other parameters are set as $t = J = S = a = 1$, $\Delta = M = 0.3$, $J_{0} = 0.4$, $K=0.1$, and $\epsilon = 1/3$.
    }
    \label{fig:strain_B1g_anisotropy_attribution}
\end{figure}

For the type-II magnet, $\chi^{B_{1g}}_{\rm sf}$ vanishes identically.
As shown in Appendix~\ref{app:b1g_vanishing}, the reflection $(k_x,k_y)\mapsto(k_y,k_x)$ is a symmetry of the type-II model that acts on neither the spin nor the sublattice index, so that the contribution of each sublattice and each magnon mode is isotropic by itself.
The whole response, and not merely its slope, therefore vanishes at every $\epsilon$, and the type-II curve accordingly vanishes at all frequencies in Fig.~\ref{fig:chi_b1g_frequency}.
In the type-I and type-III magnets the momentum-dependent term $d_{z}(\vv{k})\tau_{z}$ forces the same reflection to interchange the two sublattices, so that the two sublattice contributions cancel in the unperturbed sum but add in the $\eta_{\alpha}$-weighted combination generated by the strain.
A finite $\chi^{B_{1g}}_{\rm sf}$ thus requires the sublattice-resolved response to be anisotropic by itself, which occurs only when $t'$ or $\delta J$ is finite.
Because that response is the $B_{1g}$ anisotropy of a single sublattice, it is linear in $\delta_{\rm ani}$, and so is $\chi^{B_{1g}}_{\rm sf}$; the linearity is visible at small $\delta_{\rm ani}$ in Fig.~\ref{fig:strain_B1g_anisotropy_attribution}.
The sublattice-selective perturbation therefore promotes the altermagnetic signature to first order in $\delta_{\rm ani}$, whereas the unperturbed spin-flip channel distinguishes the type-III magnet from the type-II one only at second order.

To end this section, we examine how the electronic and the magnonic anisotropies contribute to the $B_{1g}$ strain susceptibility.
Figure~\ref{fig:strain_B1g_anisotropy_attribution} shows $\chi^{B_{1g}}_{\rm sf}$ as a function of $\delta_{\rm ani}$ at $\omega=0.03$ in (i) and at $\omega=1.3$ in (ii).
The blue and orange curves correspond to the electronic and magnonic anisotropy contributions, which are obtained by setting $(t',\delta J)$ to $(\delta_{\rm ani},0)$ and $(0,\delta_{\rm ani})$, respectively, and the black curve corresponds to the full anisotropic case $(t',\delta J)=(\delta_{\rm ani},\delta_{\rm ani})$, which is the type-III magnet.
The green curve shows the type-I magnet for comparison; since its magnon branches carry no $d$-wave splitting, only $t'$ is varied there.
In every panel the electronic contribution dominates, whereas the magnonic one stays small; it is negative at $\omega=0.03$, and at $\omega=1.3$ it is almost negligible and even changes sign between the two temperatures.
The combined case is not the sum of the two, and the difference defines a cross contribution that appears only when $t'$ and $\delta J$ are both finite.
This cross contribution is positive at $\omega=0.03$, where it grows with temperature and makes the combined susceptibility exceed the electronic one at $T=1.0$, whereas it is negative at $\omega=1.3$, where the combined susceptibility stays below the electronic one at both temperatures.

The picture becomes simpler once the $B_{1g}$ component is normalized by the $A_{1g}$ one,
\begin{align}
    \sigma^{A_{1g}}_{\rm sf}=\sigma_{xx,\downarrow\uparrow}+\sigma_{yy,\downarrow\uparrow}.
\end{align}
As shown in the lower rows of Fig.~\ref{fig:strain_B1g_anisotropy_attribution}, the full anisotropic case then agrees with the electronic contribution alone to within a few percent in all four combinations of frequency and temperature.
The origin of this agreement differs between the two frequencies: at $\omega=0.03$ the magnonic and the cross contributions are individually sizable but cancel each other almost exactly, whereas at $\omega=1.3$ both are small compared with the electronic contribution.
In either case the magnonic and the cross contributions modify the isotropic and the anisotropic parts of the spin-flip conductivity in a similar manner, so that the relative anisotropy of the spin-flip conductivity is governed by the electronic anisotropy alone.
The type-I curve shows how differently the two quantities separate the magnets.
Both magnets are linear in the anisotropy at small $t'$, as the symmetry argument above requires, so that the separation there is one of magnitude alone: in $\chi^{B_{1g}}_{\rm sf}$ itself the type-I magnet is smaller than the type-III one by more than two orders of magnitude at $\omega=0.03$ and $t'=0.2$.
The type-I susceptibility nevertheless grows rapidly once $t'\gtrsim0.3$, reaching a quarter of the type-III value at $t'=0.5$ and $T=0.2$ and seventy percent of it at $T=1.0$, and at $\omega=1.3$ the two are already comparable at $t'=0.2$, consistent with the crossing seen in Fig.~\ref{fig:chi_b1g_frequency}.
As shown in Appendix~\ref{app:type_i_growth}, this growth is a resonance effect: the weight of the momentum-resolved integrand is concentrated in a narrow shell of magnon momenta, and it is the approach of the composite electron--magnon resonance to the measuring frequency, rather than the deformation of the Fermi surface, that makes it grow.
Once normalized by the $A_{1g}$ component, by contrast, the type-I curve lies close to the type-III one everywhere, reaching three quarters of it at $t'=0.2$ and coinciding with it at $t'=0.5$.
The normalized ratio therefore measures the anisotropy of the spin-flip channel but does not by itself separate the two magnets; that separation is carried by $\chi^{B_{1g}}_{\rm sf}$ itself, and only at small anisotropy and low frequency.

\section{Conclusion}
We have computed the magnon-mediated charge-spin conversion matrix for the type-I (ferromagnetic), type-II (antiferromagnetic), and type-III (altermagnetic) magnets within a common four-band model, treating the spin-diagonal channel at one-loop order with the magnon self-energy included and the spin-flip channel at leading order in the electron-magnon coupling.
The diagonal conversion $\sigma^{\rm e-s}_{\mu\mu}$ is isotropic in the type-I magnet, vanishes in the type-II magnet, and is anisotropic in the type-III magnet; at leading one-rung order, the spin-flip channel drops out because $\Phi_{\mu\mu,\uparrow\downarrow}^{\rm R}=\Phi_{\mu\mu,\downarrow\uparrow}^{\rm R}$.
Decomposing its $B_{1g}$ anisotropy into an electronic source $t'$ and a magnonic source $\delta J$ shows that the former dominates; hence the anisotropic charge-spin conversion originates in the itinerant spin splitting that underlies the spin-splitter effect rather than in the magnons.

The magnon-mediated spin drag itself, isolated as $\sigma^{\rm e-e}_{\mu\nu}-\sigma^{\rm s-s}_{\mu\nu} = 2\lb\sigma_{\mu\nu,\up\down}+\sigma_{\mu\nu,\down\up}\rb$, behaves almost identically in the type-II and type-III magnets.
Their near agreement reflects the symmetry-enforced absence of a linear anisotropy: the $C_{4}$ or spin-flip $C_{4}$ symmetry, together with $\Phi_{\mu\mu,\uparrow\downarrow}^{\rm R}=\Phi_{\mu\mu,\downarrow\uparrow}^{\rm R}$, makes the spin-flip conductivity $C_{4}$-even, leaving differences only at second and higher orders in $\delta_{\rm ani}$.
In the spin-orbit-coupling-free limit, where the spin point group applies, the magnon-mediated spin-flip channel therefore carries no first-order signature of the altermagnetic order and cannot by itself distinguish an altermagnet from a conventional antiferromagnet \cite{Altermagnet_2022_Liu}.
The magnon leaves a spectroscopic trace as a high-frequency dip governed by the combined energy scale of an electronic spin flip and a magnon excitation.

What distinguishes them is the coupling contribution to the response of this channel under a perturbation that lowers the bulk $B_{1g}$ symmetry.
Modulating only the electron-magnon coupling in a sublattice-selective way, we find a $B_{1g}$ spin-flip response linear in the perturbation strength $\epsilon$, whose susceptibility $\chi^{B_{1g}}_{\rm sf}$ is finite in the type-III magnet, small in the type-I magnet, and identically zero in the type-II magnet, where the two sublattices are equivalent.
This susceptibility is itself linear in $\delta_{\rm ani}$, so that the perturbation promotes the altermagnetic signature from the second order, at which it appears in the unperturbed channel, to the first.
The vanishing in the type-II magnet is exact and follows from symmetry alone, whereas the separation from the type-I magnet is quantitative and, in the low frequency region, amounts to about two orders of magnitude.
Since a ferromagnet is identified by its net magnetization in any case, a finite $\chi^{B_{1g}}_{\rm sf}$ at low frequency is, among magnets with no net magnetization, unique to altermagnetic order, and it is in this sense that it provides a fingerprint of altermagnetic order in the coupling contribution to magnon-mediated transport.
The enhancement in the type-III magnet follows from its electronic anisotropy, which ties the spin and the sublattice degrees of freedom together.
Decomposing this susceptibility into its electronic and magnonic sources shows that the magnonic contribution is small and of the opposite sign, and that, once normalized by the $A_{1g}$ component, the relative anisotropy is governed by the electronic anisotropy alone.

As shown in Sec.~\ref{sec:parity}, for equal populations of the inverse magnetic domains $\mathcal N=\pm1$, $\sigma_{\mu\nu}^{\rm e-s}$ cancels because it is odd in $\mathcal N$, whereas $\sigma_{\mu\nu}^{\rm e-e}-\sigma_{\mu\nu}^{\rm s-s}$ and $\chi_{\rm sf}^{B_{1g}}$ remain because they are even.
The strain-induced $B_{1g}$ response therefore does not require a single magnetic-domain sample, although it cancels upon averaging over equally populated $\pi/2$-rotated twins.
The spin-dependent electron chemical potential is well defined only on time scales shorter than spin relaxation.
Accordingly, the spin-response results apply at frequencies above the electron spin-relaxation rate.
Extending the leading-order treatment to a conserving approximation that retains the vertex corrections and the self-energy insertions of the same order, evaluating the response that a physical strain generates through the deformed electron and magnon dispersions, and translating $\chi^{B_{1g}}_{\rm sf}$ into a concrete experimental geometry and physical units, are left for future work.

\begin{acknowledgments}
    The author is grateful to Ai Yamakage for insightful discussions and valuable suggestions throughout this study.
    This work was supported by JST SPRING (Grant No.~JPMJSP2125).
\end{acknowledgments}

\appendix
\input{appendix.tex}

\bibliography{ref}
\end{document}

%% file: appendix.tex
\section{Ferromagnetic magnon}\label{app:FM_magnon}
To consider the ferromagnetic magnon excitations, we apply the Holstein-Primakoff transformation with spin-wave approximation that $S_{\vv{r}_{\alpha},\alpha}^{+} = \sqrt{2S} a_{\vv{r}_{\alpha},\alpha}$,  $S_{\vv{r}_{\alpha},\alpha}^{-} = \sqrt{2S} a_{\vv{r}_{\alpha},\alpha}^\dagger$ and $S_{\vv{r}_{\alpha},\alpha}^{z} = S - a_{\vv{r}_{\alpha},\alpha}^\dagger a_{\vv{r}_{\alpha},\alpha}$, where $S_{\vv{r}_{\alpha},\alpha}^{\pm} = S_{\vv{r}_{\alpha},\alpha}^{x} \pm i S_{\vv{r}_{\alpha},\alpha}^{y}$.
The operator $a_{\vv{r}_{\alpha},\alpha}$ is the magnon annihilation operator at position $\vv{r}_{\alpha}$ on sublattice $\alpha$.
After Fourier transformation $a_{\vv{r}_{\alpha},\alpha} = \frac{1}{\sqrt{N}}\sum_{\vv{q}} a_{\vv{q}\alpha} e^{-i\vv{q}\cdot\vv{r}_{\alpha}}$, the magnon Hamiltonian is written as
\begin{align}
    \mathcal{H}_{\rm I}^{\rm mag}
    ={}&\sum_{\vv{q}}
    \mqty(a_{\vv{q}A}^\dagger & a_{\vv{q}B}^\dagger)
    \mqty(
        \gamma_{\vv{q}} &   \gamma_{0,\vv{q}}^{\rm I} \\
        \gamma_{0,\vv{q}}^{\rm I} & \gamma_{\vv{q}}
    )
    \mqty(a_{\vv{q}A} \\ a_{\vv{q}B}),
\end{align}
where $\gamma_{\vv{q}} = (4J_{0} + 2K)S + JS \sum_{\mu} \left(1- \cos(a q_{\mu})\right)$ and $\gamma_{0,\vv{q}}^{\rm I} = -J_{0}S\Gamma_{\vv{q}}=-4J_{0}S\cos(aq_x/2)\cos(aq_y/2)$.
By the unitary transformation that 
\begin{align}
    \mqty(a_{\vv{q}A} \\ a_{\vv{q}B})
    = \frac{1}{\sqrt{2}}
    \mqty(1 & 1 \\ 1 & -1)
    \mqty(a_{\vv{q}+} \\ a_{\vv{q}-}),
\end{align}
we can diagonalize the magnon Hamiltonian as
\begin{align}
    \mathcal{H}_{\rm I}^{\rm mag}
    =\sum_{\vv{q},m=\pm}
    \left(\gamma_{\vv{q}} + m \gamma_{0,\vv{q}}^{\rm I}\right)
    a_{\vv{q}m}^\dagger a_{\vv{q}m}.
\end{align}

\section{Antiferromagnetic magnon}\label{app:AF_magnon}
Now, we introduce the Holstein-Primakoff transformation around the collinear antiferromagnetic ground state with spin-wave approximation that $S_{\vv{r}_A,A}^{+} = \sqrt{2S}b_{\vv{r}_A,A}$, $S_{\vv{r}_A,A}^{-} = \sqrt{2S}b_{\vv{r}_A,A}^{\dagger}$, $S_{\vv{r}_A,A}^{z} = S - b_{\vv{r}_A,A}^{\dagger}b_{\vv{r}_A,A}$ for the $A$-sublattice.
For the $B$-sublattice, we have $S_{\vv{r}_B,B}^{+} = \sqrt{2S}b_{\vv{r}_B,B}^{\dagger}$, $S_{\vv{r}_B,B}^{-} = \sqrt{2S}b_{\vv{r}_B,B}$, and $S_{\vv{r}_B,B}^{z} = -S + b_{\vv{r}_B,B}^{\dagger}b_{\vv{r}_B,B}$, where $b_{\vv{r}_{\alpha},\alpha}$ is the magnon annihilation operator at position $\vv{r}_{\alpha}$ on sublattice $\alpha$.
By using the Fourier transformation $b_{\vv{r}_{\alpha},\alpha} = \frac{1}{\sqrt{N}}\sum_{\vv{q}} b_{\vv{q}\alpha} e^{-i\vv{q}\cdot\vv{r}_{\alpha}}$, the magnon Hamiltonian is written as
\begin{align}
    \mathcal{H}_{\rm AF}^{\rm mag}
    ={}&\sum_{\vv{q}}
    \mqty(b_{\vv{q}A}^\dagger & b_{-\vv{q}B})
    \mqty(
        \gamma_{\vv{q}}^A & \gamma_{0,\vv{q}}^{\rm AF} \\
        \gamma_{0,\vv{q}}^{\rm AF} & \gamma_{\vv{q}}^B
    )
    \mqty(b_{\vv{q}A} \\ b_{-\vv{q}B}^\dagger),
    \\
    \gamma_{\vv{q}}^{A}
    ={}& \gamma_{\vv{q}} - S\delta J \lb \cos(a q_x) - \cos(a q_y) \rb,
    \\
    \gamma_{\vv{q}}^{B}
    ={}& \gamma_{\vv{q}} + S\delta J \lb \cos(a q_x) - \cos(a q_y) \rb,
    \\
    \gamma_{0,\vv{q}}^{\rm AF}
    ={}& J_{0}S\Gamma_{\vv{q}}
    =4J_{0}S\cos(aq_x/2)\cos(aq_y/2),
\end{align}
where we ignored the additive normal-ordering constant, and $\gamma_{\vv{q}} = 4J_{0}S + 2KS + JS \sum_{\mu} \left(1- \cos(a q_{\mu})\right)$.
By using the Bogoliubov transformation $b_{\vv{q}A} = u_{\vv{q}} \hat{\alpha}_{\vv{q}} + v_{\vv{q}} \hat{\beta}_{\vv{q}}^{\dagger}$ and $b_{-\vv{q}B}^{\dag} = v_{\vv{q}} \hat{\alpha}_{\vv{q}} + u_{\vv{q}} \hat{\beta}_{\vv{q}}^{\dagger}$ with $u_{\vv{q}}^2 - v_{\vv{q}}^2 = 1$, we can diagonalize the magnon Hamiltonian as
\begin{align}
    \mathcal{H}_{\rm AF}^{\rm mag}
    &=\sum_{\vv{q}}
    \left(\omega_{\vv{q}}^{+}\hat{\alpha}_{\vv{q}}^\dagger \hat{\alpha}_{\vv{q}}
    +\omega_{\vv{q}}^{-}\hat{\beta}_{\vv{q}}^\dagger \hat{\beta}_{\vv{q}}\right),
    \\
    \omega_{\vv{q}}^{\pm}
    &= E_{\vv{q}} \mp S\delta J \lb \cos(a q_x) - \cos(a q_y) \rb,
    \\
    E_{\vv{q}} &= \sqrt{\gamma_{\vv{q}}^2 - (\gamma_{0,\vv{q}}^{\rm AF})^2}  > 0,
    \quad
    u_{\vv{q}} = \sqrt{\frac{\gamma_{\vv{q}} + E_{\vv{q}}}{2E_{\vv{q}}}},
    \notag\\
    v_{\vv{q}} &= -\sgn(\gamma_{0,\vv{q}}^{\rm AF})\sqrt{\frac{\gamma_{\vv{q}} - E_{\vv{q}}}{2E_{\vv{q}}}}.
\end{align}

The leading-order Holstein--Primakoff expansion used above is controlled as long as the magnon population stays small compared with $2S$.
For the parameters used throughout this paper, the easy-axis anisotropy opens the magnon gap $E_{\vv{q}=0}=S\sqrt{2K(8J_{0}+2K)}\simeq0.82$, which keeps the population modest even at the highest temperature considered.
Evaluating
\begin{align}
    \Lb b_{A}^{\dagger}b_{A}\Rb
    =\frac{1}{N}\sum_{\vv{q}}
    \LB
        u_{\vv{q}}^{2}\,n(\omega_{\vv{q}}^{+})
        +v_{\vv{q}}^{2}\lb 1+n(\omega_{\vv{q}}^{-})\rb
    \RB
\end{align}
over the Brillouin zone gives $0.038$ at $T=0.2$ and $0.123$ at $T=1.0$, so that the sublattice magnetization $\Lb S^{z}\Rb=S-\Lb b_{A}^{\dagger}b_{A}\Rb$ is reduced by only $4\%$ and $12\%$, respectively; the type-II and type-III values agree to within $1\%$, and the corresponding numbers for the ferromagnetic magnons of Appendix~\ref{app:FM_magnon} are $0.005$ and $0.118$.
The linear spin-wave description is therefore well controlled at the temperatures used here.

\section{Electron self-energy due to the electron--magnon interaction}\label{app:self_energy}

\begin{figure}
    \centering 
    \includegraphics[scale=0.7]{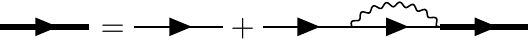}
    \caption{
        Feynman diagram for the Dyson equation of the electron Green function.
        The solid line is the electron propagator, and the wavy line is the magnon propagator.
    }
    \label{fig:Dyson-eq}
\end{figure}

In this section, we derive the electron self-energy due to the electron--magnon interaction.
The self-energy in the Dyson equation, shown diagrammatically in Fig.~\ref{fig:Dyson-eq}, is given by the Fock diagram as
\begin{align}
    \Sigma_{\alpha,\sigma}(\vv{k},i\epsilon_n; \mathcal{N})
    &=
    -\frac{1}{\beta_{T}N}\sum_{\vv{q},m,\nu_\ell}
    C_{\alpha, m}(\vv{q})
    D_{m,\mathcal{N}}(\vv{q},i\nu_\ell)
    \notag\\
    &\quad\times
    G_{\alpha,\sigma'}^{(0)}(\vv{k}+\sigma\vv{q},i\epsilon_n+i\sigma\nu_\ell;\mathcal{N}),
\end{align}
where $C_{\alpha, m}(\vv{q})$ is the magnon-electron interaction vertex, which is given as $C_{\alpha, m}(\vv{q}) = \Delta^2/S$ for the type-I magnet, and for the type-II and type-III magnets, it is given as $C_{\alpha, m}(\vv{q}) = |u_{\vv{q}}|^2 (2M^2/S)$ for $(m,\alpha) = (+,A), (-,B)$ and $C_{\alpha, m}(\vv{q}) = |v_{\vv{q}}|^2 (2M^2/S)$ for $(m,\alpha) = (-,A), (+,B)$.
The spin index $\sigma' = -\sigma$ is the spin opposite to $\sigma$.
By taking the summation over the Matsubara frequency $i\nu_\ell$, we can obtain the self-energy as
\begin{align}
    \Sigma_{\alpha,\sigma}(\vv{k},i\epsilon_n; \mathcal{N})
    &=
    \frac{1}{N}\sum_{\vv{q},m}
    C_{\alpha, m}(\vv{q})\zeta_{m,\mathcal{N}}
    \notag\\
    &\times
    \frac{
        f(\sigma \xi_{\vv{k}+\sigma\vv{q},\alpha,\sigma'}) + n(\zeta_{m,\mathcal{N}}\omega_{m,\vv{q}})
    }{
        i\epsilon_n - \xi_{\vv{k}+\sigma\vv{q},\alpha,\sigma'} + \sigma \zeta_{m,\mathcal{N}}\omega_{m,\vv{q}}
    },
\end{align}
where we use the relationship for the Fermi and Bose distribution functions as $f(-x) = 1 - f(x)$ and $n(-x) = -[1 + n(x)]$.
By analytically continuing $i\epsilon_n \to \epsilon + i0$, we can obtain the retarded self-energy $\Sigma_{\alpha,\sigma}^{\rm R}(\vv{k},\epsilon; \mathcal{N})$.
From the Dyson equation, the retarded Green function is given by
\begin{align}
    G_{\alpha,\sigma}^{\rm R}(\vv{k},\epsilon; \mathcal{N})
    = \frac{1}{\epsilon + i0^{+} - \xi_{\vv{k}\alpha\sigma} - \Sigma_{\alpha,\sigma}^{\rm R}(\vv{k},\epsilon; \mathcal{N})}.
\end{align}

\section{Derivation of spin-diagonal conductivity}\label{app:spin_diagonal_conductivity}
In this section, we derive the spin-diagonal conductivity $\sigma_{\mu\nu,\sigma\sigma}$ by using the one-loop diagram.
We start from the current-current correlation function in the Matsubara formalism as
\begin{align}
    \Phi_{\mu\nu,\sigma\sigma}(i\Omega_{\lambda}; \mathcal{N})
    &=
    -\frac{1}{\beta_{T}N}\sum_{\vv{k},\epsilon_n,\alpha}
    j_{\mu,\alpha\sigma}(\vv{k})
    j_{\nu,\alpha\sigma}(\vv{k})
    \notag\\
    &\times
    G_{\alpha,\sigma}(\vv{k},i\epsilon_n; \mathcal{N})
    G_{\alpha,\sigma}(\vv{k},i\epsilon_n+i\Omega_{\lambda}; \mathcal{N}),
\end{align}
where $j_{\mu,\alpha\sigma}(\vv{k}) = -e\pdv{\xi_{\vv{k}\alpha\sigma}}{k_\mu}$ is the current vertex.
To perform the Matsubara summation over $i\epsilon_n$, we use the spectral representation of the Green function as
\begin{align}
    G_{\alpha,\sigma}(\vv{k},i\epsilon_n; \mathcal{N})
    = \int_{-\infty}^{\infty} d\epsilon
    \frac{A_{\alpha,\sigma}(\vv{k},\epsilon; \mathcal{N})}{i\epsilon_n - \epsilon},
\end{align}
where $A_{\alpha,\sigma}(\vv{k},\epsilon; \mathcal{N}) = -\frac{1}{\pi}\mathrm{Im} G_{\alpha,\sigma}^{\rm R}(\vv{k},\epsilon; \mathcal{N})$ is the spectral function, which is normalized as $\int d\epsilon\, A_{\alpha,\sigma}(\vv{k},\epsilon; \mathcal{N}) = 1$ consistently with the spectral representation above.
The Matsubara summation can be performed as
\begin{align}
    -\frac{1}{\beta_{T}}\sum_{\epsilon_n}
    \frac{1}{i\epsilon_n - \epsilon}
    \frac{1}{i\epsilon_n + i\Omega_{\lambda} - \epsilon'}
    = -\frac{f(\epsilon) - f(\epsilon')}{i\Omega_{\lambda} - \epsilon' + \epsilon},
\end{align}
where $f(\epsilon) = 1/(e^{\beta_{T}\epsilon} + 1)$ is the Fermi distribution function.
Thus, the current-current correlation function is written as
\begin{align}
    \Phi_{\mu\nu,\sigma\sigma}(i\Omega_{\lambda}; \mathcal{N})
    &=
    -\frac{1}{N}\sum_{\vv{k},\alpha}
    j_{\mu,\alpha\sigma}(\vv{k})
    j_{\nu,\alpha\sigma}(\vv{k})
    \notag\\
    &\times 
    \int d\epsilon d\epsilon' 
    A_{\alpha,\sigma}(\vv{k},\epsilon; \mathcal{N})
    A_{\alpha,\sigma}(\vv{k},\epsilon'; \mathcal{N})
    \notag\\ 
    &\times
    \frac{f(\epsilon) - f(\epsilon')}{i\Omega_{\lambda} - \epsilon' + \epsilon}.
\end{align}
Here, we use the analytic continuation $i\Omega_{\lambda} \to \omega + i0$ to obtain the retarded correlation function and apply $\mathrm{Im} (x + i0^+ )^{-1} = -\pi \delta(x)$.
The imaginary part of the retarded correlation function is written as
\begin{align}
    \mathrm{Im} \Phi_{\mu\nu,\sigma\sigma}^{\rm R}(\omega; \mathcal{N})
    &=
    \frac{\pi}{N}\sum_{\vv{k},\alpha}
    j_{\mu,\alpha\sigma}(\vv{k})
    j_{\nu,\alpha\sigma}(\vv{k})
    \notag\\
    &\times
    \int d\epsilon
    A_{\alpha,\sigma}(\vv{k},\epsilon; \mathcal{N})
    A_{\alpha,\sigma}(\vv{k},\epsilon+\omega; \mathcal{N})
    \notag\\
    &\times
    \left[f(\epsilon) - f(\epsilon+\omega)\right].
\end{align}
Finally, we can obtain the spin-diagonal conductivity from the Kubo formula as
\begin{align}
    \mathrm{Re}\sigma_{\mu\nu,\sigma\sigma}(\omega; \mathcal{N})
    &=
    \frac{
        \mathrm{Im} \Phi_{\mu\nu,\sigma\sigma}^{\rm R}(\omega; \mathcal{N})
        }{\omega},
\end{align}
where $\mathrm{Re}\sigma_{\mu\nu,\sigma\sigma}(\omega; \mathcal{N})$ is the regular dissipative part of the spin-diagonal conductivity, which does not include the Drude/contact term $\propto\delta(\omega)$.

\section{Derivation of spin-flip conductivity}\label{app:spin_flip_conductivity}
In this section, we derive the spin-flip conductivity $\sigma_{\mu\nu,\sigma\sigma'}$ by using the one-rung diagram.
We start from the current-current correlation function in the Matsubara formalism as
\begin{align}
    &\Phi_{\mu\nu,\sigma\sigma'}(i\Omega_{\lambda}; \mathcal{N})
    \notag\\
    &=
    \frac{1}{\beta_{T}^2N^2}\sum_{\vv{k},\vv{q},\epsilon_n,\nu_\ell,\alpha,m}
    j_{\mu,\alpha\sigma}(\vv{k})
    j_{\nu,\alpha\sigma'}(\vv{k}+\sigma\vv{q})
    \notag\\
    &\times
    C_{\alpha, m}(\vv{q})
    \notag\\ 
    &\times
    G_{\alpha,\sigma}^{(0)}(\vv{k},i\epsilon_n;\mathcal{N})
    G_{\alpha,\sigma}^{(0)}(\vv{k},i\epsilon_n+i\Omega_{\lambda};\mathcal{N})
    \notag\\
    &\times
    G_{\alpha,\sigma'}^{(0)}(\vv{k}+\sigma\vv{q},i\epsilon_n+i\sigma\nu_\ell;\mathcal{N})
    D_{m,\mathcal{N}}(\vv{q},i\nu_\ell)
    \notag\\ 
    &\times
    G_{\alpha,\sigma'}^{(0)}(\vv{k}+\sigma\vv{q},i\epsilon_n+i\sigma\nu_\ell+i\Omega_{\lambda};\mathcal{N}),
\end{align}
where $G_{\alpha,\sigma}^{(0)}(\vv{k},i\epsilon_n;\mathcal{N}) = 1/[i\epsilon_n - \xi_{\vv{k}\alpha\sigma}(\mathcal{N})]$ is the noninteracting electron Green's function, and $D_{m,\mathcal{N}}(\vv{q},i\nu_\ell) = \zeta_{m,\mathcal{N}}/ (i\nu_\ell - \zeta_{m,\mathcal{N}}\omega_{m,\vv{q}})$ is the magnon propagator.
First, we perform the Matsubara summation over $i\epsilon_n$ as 
\begin{align}
    &\frac{1}{\beta_{T}}\sum_{\epsilon_n}
    \frac{1}{i\epsilon_n - a}\frac{1}{i\epsilon_n + i\Omega_\lambda -a}
    \notag\\ 
    &\quad\times
    \frac{1}{i\epsilon_n + i\sigma\nu_{\ell} - b} 
    \frac{1}{i\epsilon_n + i\sigma\nu_{\ell} + i\Omega_\lambda - b}
    \notag\\
    &=
    \frac{2(f(b) - f(a))}
    {i\sigma\nu_{\ell} - \Delta E}
    \frac{1}{(i\sigma \nu_{\ell} - \Delta E)^2 - (i\Omega_\lambda)^2}
    \notag\\ 
    &=: \mathcal{F}(a,b;i\sigma\nu_\ell, i\Omega_\lambda),
\end{align}
where $\Delta E = b - a$.
Using this result, we can perform the Matsubara summation over $i\nu_\ell$ as
\begin{align}
    &\mathcal{K}_{\sigma,m,\mathcal{N}}(a,b;i\Omega_\lambda)
    \notag\\
    &=
    \frac{1}{\beta_{T}}\sum_{\nu_\ell}
    D_{m,\mathcal{N}}(\vv{q},i\nu_\ell)
    \mathcal{F}(a,b;i\sigma\nu_\ell, i\Omega_\lambda)
    \notag\\
    &=
    \frac{
        2\sigma\zeta_{m, \mathcal{N}}
        (f(b) - f(a))
        (n(\zeta_{m, \mathcal{N}}\omega_{m,\vv{q}}) - n(\sigma\Delta E))
    }
    {
        (\sigma\Delta E - \zeta_{m, \mathcal{N}}\omega_{m,\vv{q}})
        \left\{ 
            (\sigma\Delta E - \zeta_{m, \mathcal{N}}\omega_{m,\vv{q}})^2 - (i\Omega_\lambda)^2
        \right\}
    }.
\end{align}
Thus, we obtain the kernel function $\mathcal{K}_{\sigma,m,\mathcal{N}}(a,b;i\Omega_\lambda)$, which is the key quantity to calculate the spin-flip conductivity.
Using this kernel function, the current-current correlation function is written as
\begin{align}
    \Phi_{\mu\nu,\sigma\sigma'}(i\Omega_{\lambda}; \mathcal{N})
    &=
    \frac{1}{N^2}\sum_{\vv{k},\vv{q},\alpha,m}
    j_{\mu,\alpha\sigma}(\vv{k})
    j_{\nu,\alpha\sigma'}(\vv{k}+\sigma\vv{q})
    \notag\\
    \times&
    C_{\alpha, m}(\vv{q})
    \mathcal{K}_{\sigma,m,\mathcal{N}}(\xi_{\vv{k}\alpha\sigma},\xi_{\vv{k}+\sigma\vv{q},\alpha\sigma'};i\Omega_\lambda).
\end{align}
Thus the spin-flip conductivity is obtained from the Kubo formula as
\begin{align}
    \mathrm{Re}\sigma_{\mu\nu,\sigma\sigma'}(\omega; \mathcal{N})
    &=
    \frac{
        \mathrm{Im} \Phi_{\mu\nu,\sigma\sigma'}^{\rm R}(\omega; \mathcal{N})
        }{\omega},
\end{align}
where retarded correlation function $\Phi_{\mu\nu,\sigma\sigma'}^{\rm R}(\omega; \mathcal{N})$ is obtained by the analytic continuation $i\Omega_\lambda \to \omega + i0^+$.

\section{Vanishing of the $B_{1g}$ strain susceptibility in the type-II magnet}\label{app:b1g_vanishing}
In this section, we show that the $B_{1g}$ strain susceptibility of the spin-flip conductivity vanishes identically in the type-II magnet, and we identify what distinguishes the type-I and type-III magnets from it.

We first resolve the response into the contributions of each sublattice $\alpha$ and magnon mode $m$.
The one-rung correlation function derived in Appendix~\ref{app:spin_flip_conductivity} is linear in the vertex weight $C_{\alpha,m}(\vv{q})$, so the coupling modulation of Eq.~\eqref{eq:coupling_modulation} gives
\begin{align}
    \sigma_{\rm sf}^{B_{1g}}(\epsilon)
    &=\sum_{\alpha,m}\lb 1+\eta_{\alpha}\epsilon\rb X_{\alpha m},
    \notag\\
    X_{\alpha m}
    &\equiv
    \sigma_{xx,\downarrow\uparrow}^{(\alpha m)}
    -\sigma_{yy,\downarrow\uparrow}^{(\alpha m)},
    \label{eq:b1g_decomposition}
\end{align}
where $\sigma_{\mu\mu,\downarrow\uparrow}^{(\alpha m)}$ is the contribution of sublattice $\alpha$ and mode $m$ evaluated at $\epsilon=0$.

Consider now the reflection $R:(k_x,k_y)\mapsto(k_y,k_x)$, which acts on neither the spin nor the sublattice index.
In $H_{\rm II}$ the exchange term $M\sigma_{z}\otimes\tau_{z}$ carries no momentum dependence and $\epsilon_{0}(\vv{k})$ is invariant under $R$, so that
\begin{align}
    \xi_{R\vv{k},\alpha\sigma}=\xi_{\vv{k}\alpha\sigma}
    \label{eq:xi_R_invariance}
\end{align}
holds for each sublattice and each spin separately.
Setting $\delta J=0$ leaves the magnon energies $\omega_{m,\vv{q}}=E_{\vv{q}}$ and the vertex weights $C_{\alpha,m}(\vv{q})$, which depend on $\vv{q}$ only through $\gamma_{\vv{q}}$ and $\Gamma_{\vv{q}}$, invariant under $R$ as well, and hence so is the kernel $\mathcal{K}_{\sigma,m,\mathcal{N}}$.
Differentiating Eq.~\eqref{eq:xi_R_invariance} with respect to $k_{x}$ shows that the two current vertices are interchanged,
\begin{align}
    j_{x,\alpha\sigma}(R\vv{k})=j_{y,\alpha\sigma}(\vv{k}).
\end{align}
Since $R$ is linear, $R\vv{k}+\sigma R\vv{q}=R(\vv{k}+\sigma\vv{q})$, and the Brillouin zone is invariant under $R$.
Changing the summation variables as $\vv{k}\to R\vv{k}$ and $\vv{q}\to R\vv{q}$ therefore leaves every factor of the one-rung expression unchanged except for the current vertices, which are exchanged, and we obtain
$\sigma_{xx,\downarrow\uparrow}^{(\alpha m)}=\sigma_{yy,\downarrow\uparrow}^{(\alpha m)}$, that is, $X_{\alpha m}=0$ for every $\alpha$ and $m$ separately.
By Eq.~\eqref{eq:b1g_decomposition} this gives not merely a vanishing slope but $\sigma_{\rm sf}^{B_{1g}}(\epsilon)=0$ at every $\epsilon$.

The same decomposition shows why the type-I and type-III magnets respond at all.
There $d_{z}(\vv{k})$ changes sign under $R$, so the reflection must be accompanied by the interchange of the two sublattices, and in the type-III magnet also by a spin flip and by $m\to-m$; writing $\bar{\alpha}$ for the sublattice opposite to $\alpha$,
\begin{align}
    \xi_{R\vv{k},\alpha\sigma}
    =\begin{cases}
        \xi_{\vv{k}\bar{\alpha}\sigma}, & \text{type I},\\
        \xi_{\vv{k}\bar{\alpha}\bar{\sigma}}, & \text{type III}.
    \end{cases}
\end{align}
The reflection then relates the two sublattices instead of acting within each of them, and, using $\Phi_{\mu\mu,\uparrow\downarrow}^{\rm R}=\Phi_{\mu\mu,\downarrow\uparrow}^{\rm R}$ for the type-III magnet, we find $X_{Am}=-X_{Bm}$ for the type-I magnet and $X_{Am}=-X_{B,-m}$ for the type-III magnet.
The unweighted sum therefore vanishes, $\sigma_{\rm sf}^{B_{1g}}(0)=\sum_{\alpha,m}X_{\alpha m}=0$, whereas the $\eta_{\alpha}$-weighted sum survives,
\begin{align}
    \chi^{B_{1g}}_{\rm sf}
    =\sum_{\alpha,m}\eta_{\alpha}X_{\alpha m}
    =2\sum_{m}X_{Am}.
    \label{eq:chi_weighted_sum}
\end{align}

\section{Growth of the type-I susceptibility with the electronic anisotropy}\label{app:type_i_growth}
In this section we trace the rapid growth of $\chi^{B_{1g}}_{\rm sf}$ of the type-I magnet for $t'\gtrsim0.3$, seen in Fig.~\ref{fig:strain_B1g_anisotropy_attribution}, to the composite electron--magnon resonance.

We resolve the susceptibility in the magnon momentum by introducing the $\vv{k}$-summed integrand of the $B_{1g}$ component,
\begin{align}
    &G^{B_{1g}}_{\alpha,m}(\omega,\vv{q})
    \notag\\
    ={}&\frac{1}{N}\sum_{\vv{k}}
    \LB
        j_{x,\alpha\up}(\vv{k})\,j_{x,\alpha\down}(\vv{k}+\vv{q})
        -j_{y,\alpha\up}(\vv{k})\,j_{y,\alpha\down}(\vv{k}+\vv{q})
    \RB
    \notag\\
    &\times
    \mathcal{K}_{\up,m,\mathcal{N}}\lb\xi_{\vv{k}\alpha\up},\xi_{\vv{k}+\vv{q},\alpha\down};\omega+i\eta\rb .
\end{align}
With $X_{Am}=-X_{Bm}$ from Eq.~\eqref{eq:chi_weighted_sum}, the susceptibility is carried by a single sublattice,
\begin{align}
    \chi^{B_{1g}}_{\rm sf}(\omega)
    =\frac{2}{\omega N}\,\mathrm{Im}\sum_{\vv{q},m}
    C_{A,m}(\vv{q})\,G^{B_{1g}}_{A,m}(\omega,\vv{q}).
    \label{eq:chi_q_resolved}
\end{align}
For the type-I magnet $C_{A,m}=\Delta^{2}/S$ is independent of $\vv{q}$ and of $m$, and the optical branch, whose gap $(8J_{0}+2K)S$ is thermally inaccessible at the temperatures used here, contributes about three orders of magnitude less than the acoustic one; we therefore show only $m=+$.

\begin{figure}
    \centering
    \includegraphics{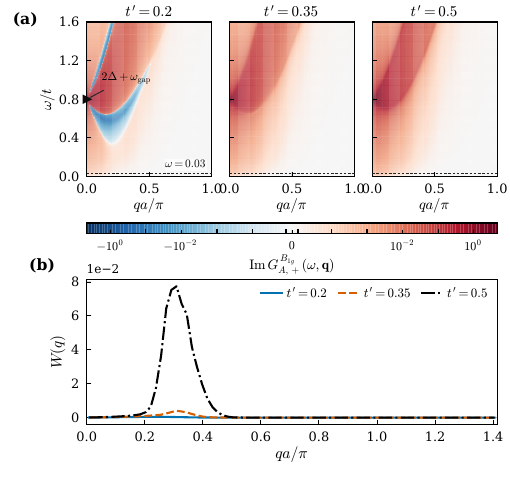}
    \caption{
        Momentum-resolved origin of the $B_{1g}$ strain susceptibility of the type-I magnet.
        (a) $\mathrm{Im}\,G^{B_{1g}}_{A,+}(\omega,\vv{q})$ along $\Gamma$--X for $t'=0.2$, $0.35$, and $0.5$, drawn on a common colour scale.
        The arrow marks the composite energy $2|\Delta|+\omega_{\rm gap}=0.8$ at $\vv{q}=0$, and the dashed line the frequency $\omega=0.03$ at which the susceptibility is evaluated.
        (b) The radial density $W(q)$ of Eq.~\eqref{eq:radial_density} at $\omega=0.03$, whose area equals $\chi^{B_{1g}}_{\rm sf}$.
        The optical magnon branch contributes about three orders of magnitude less and is not shown.
        In this figure, we set the magnetic domain $\mathcal{N}=+1$, temperature $T=0.2$, the chemical potential $\mu=0.5$, the broadening $\eta=0.02$, and the other parameters are set as $t=J=S=a=1$, $\Delta=0.3$, $J_{0}=0.4$, and $K=0.1$.
    }
    \label{fig:type_i_qmap}
\end{figure}

Figure~\ref{fig:type_i_qmap}(a) shows $\mathrm{Im}\,G^{B_{1g}}_{A,+}$ along $\Gamma$--X.
The weight is organized around the resonance edge that emanates from $\vv{q}=0$ at the composite energy $2|\Delta|+\omega_{\rm gap}=0.8$, the same scale that produces the high-frequency structure of Fig.~\ref{fig:chi_b1g_frequency}.
At $t'=0.2$ a region of opposite sign lies below this edge and cancels part of the positive weight; it recedes as $t'$ increases, and by $t'=0.5$ the low-frequency region carries a single sign.
To display where in the Brillouin zone the susceptibility is accumulated, we introduce the radial density
\begin{align}
    W(q)
    =\frac{2}{\omega N}\,\mathrm{Im}\sum_{m}\sum_{\vv{q}'}
    \delta\lb q-\lvert\vv{q}'\rvert\rb
    C_{A,m}(\vv{q}')\,G^{B_{1g}}_{A,m}(\omega,\vv{q}'),
    \label{eq:radial_density}
\end{align}
where $\vv{q}'$ runs over the first Brillouin zone and the delta function is evaluated as a histogram in $\lvert\vv{q}'\rvert$.
Comparison with Eq.~\eqref{eq:chi_q_resolved} shows that
\begin{align}
    \int_{0}^{\infty}dq\,W(q)=\chi^{B_{1g}}_{\rm sf}(\omega),
\end{align}
so that the area under $W(q)$ is the susceptibility itself.
Figure~\ref{fig:type_i_qmap}(b) shows $W(q)$ at the working frequency $\omega=0.03$.
The weight is concentrated in a narrow shell near $qa/\pi\simeq0.3$ and grows there by about two orders of magnitude between $t'=0.2$ and $t'=0.5$, which accounts for the growth seen in Fig.~\ref{fig:strain_B1g_anisotropy_attribution}.
The susceptibility of the type-I magnet is therefore governed by how close the composite electron--magnon resonance comes to the measuring frequency, and not by the deformation of the Fermi surface, which varies smoothly over the same range of $t'$.

\section{Numerical implementation and convergence}\label{app:numerics}
For the numerical calculations, we set $a=t=k_{\rm B}=1$.
The electron self-energy and the self-energy-dressed spin-diagonal conductivity were evaluated on uniform meshes with $N_k=N_q=24$ points per Cartesian direction over the first Brillouin zone $[-\pi/a,\pi/a)^2$.
The internal-energy integral over $\epsilon\in[-8,12]$ and the self-energy pole-energy grid over $E\in[-12,12]$ used the composite trapezoidal rule with $\Delta\epsilon=\Delta E=0.02$; for the low-frequency and anisotropy calculations, both spacings were reduced to $0.005$.
We used the retarded broadening $\eta=0.02$, with $\eta/T\leq0.1$ in all calculations.
The same $\eta$ regularizes the spin-flip response: after the Matsubara sums have been performed analytically, the continuation $i\Omega_{\lambda}\to\omega+i\eta$ replaces the sharp poles of the kernel $\mathcal{K}_{\sigma,m,\mathcal{N}}$ by Lorentzians of width $\eta$, which renders $\mathrm{Im}\,\Phi_{\mu\nu,\sigma\sigma'}^{\rm R}(\omega;\mathcal{N})$ finite.
After the Matsubara sums were performed analytically, the four-dimensional momentum integral for the spin-flip response was evaluated with a tensor-product composite Gauss--Legendre rule using two panels and 24 nodes per panel on each momentum axis, corresponding to 48 nodes per axis; the high-frequency results used 32 nodes per panel, or 64 nodes per axis.
Refining the spin-diagonal calculation to $N_k=N_q=28$ changed the combined relative $L^2$ norm by $0.418\%$, $0.397\%$, and $0.288\%$ for the type-I, type-II, and type-III magnets, respectively, while increasing the high-frequency spin-flip quadrature from 64 to 80 nodes per axis gave maximum relative changes of $0.434\%$, $0.331\%$, and $0.208\%$ for the same three magnets.
These values are below the prescribed $5\%$ convergence criterion.
For the production spin-diagonal data, the maximum spectral-sum-rule error was $4.98\times10^{-3}$, and no spectral weight outside the sampled pole-energy window was detected within numerical precision.
The $B_{1g}$ response of the type-I magnet was evaluated on four shifted uniform grids with $n=96$, whereas for the type-III magnet we used a $C_4$-symmetrized randomized quasi-Monte Carlo integration with $2^{24}$ samples, 16 repetitions, and two independent seed sets; the resulting relative standard error was $6.97\times10^{-4}$.
The same settings were used for the calculations at $T=1.0$ and at $\omega=1.3$, and for the sweeps in $\delta_{\rm ani}$ shown in Fig.~\ref{fig:strain_B1g_anisotropy_attribution}.

\section{Full frequency dependence of the spin-flip conductivity}\label{app:full_frequency}

\begin{figure}
    \centering
    \includegraphics[scale=1.0]{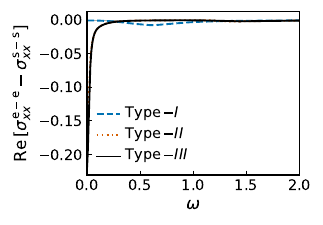}
    \caption{
        The difference between the charge--charge conductivity and the spin--spin conductivity $\mathrm{Re} \sigma^{\rm e-e}_{xx}(\omega; \mathcal{N}) - \mathrm{Re} \sigma^{\rm s-s}_{xx}(\omega; \mathcal{N})$ over the full range $0 < \omega < 2$ on a common scale, for the type-I, type-II, and type-III magnets.
        The parameters are the same as in Fig.~\ref{fig:ee_minus_ss_xx_vs_omega}.
    }
    \label{fig:ee_minus_ss_full}
\end{figure}

Figure~\ref{fig:ee_minus_ss_full} shows the real part of the difference between the charge--charge and spin--spin conductivities over the full frequency range on a common scale.
For the numerical data shown, the response is dominated by the low-frequency region below $\omega \simeq 0.1$, whose magnitude is visibly larger than that of the finite-frequency dip features discussed in Sec.~\ref{sec:spinflip}; the two windows of Fig.~\ref{fig:ee_minus_ss_xx_vs_omega} are therefore drawn on separate vertical scales.
In the interval left out between those windows, $0.3 < \omega < 0.4$, and above $\omega = 1.7$, the curves vary smoothly and show no additional structure.